\documentclass[12pt,letterpaper]{article}
\usepackage[includeheadfoot, 
            marginratio={1:1,2:3}, 
            width=413.25pt, 
            height=675pt,]{geometry}

\usepackage{amsmath}
\usepackage{amsfonts}
\usepackage{amssymb}

\usepackage{dsfont}
\usepackage{titlesec}

\numberwithin{equation}{section}

\newcommand{\eq}[1]{\begin{equation}
                     \begin{split} #1 \end{split}
                     \end{equation}}
                             
\newcommand{\ov}[1]{\overline{#1}}
\newcommand{\K}{\mathcal{K}}
\newcommand{\V}{\mathcal{V}}
\newcommand{\X}{\mathcal{X}}

\newcommand{\Q}{\mathcal{Q}}
\newcommand{\M}{\mathcal{M}}
\newcommand{\R}{\mathcal{R}}
\newcommand{\s}{\mathcal{S}}
\newcommand{\T}{\mathcal{T}}

\renewcommand{\Re}{{\rm Re}\hspace{0.5pt}}
\renewcommand{\Im}{{\rm Im}\hspace{0.5pt}}

\renewcommand{\d}{{\rm d}}

\begin{document}

\titlespacing*{\section}{0pt}{*2}{*1}
\titlespacing*{\subsection}{0pt}{*1}{*0}
\titlespacing*{\subsubsection}{0pt}{*1.5}{*0}

%%%%%%%%%%%%%%%%%%%%%%%%%%%%%%%%%%%%%%%%%%%%%%%
%%%%%%%%%%%%%%%%%%%%%%%%%%%%%%%%%%%%%%%%%%%%%%%
%%%%%%%%%%%%%%%%%%%%%%%%%%%%%%%%%%%%%%%%%%%%%%%
%%%%%%%%%%%%%%%%%%%%%%%%%%%%%%%%%%%%%%%%%%%%%%%
%%%%%%%%%%%%%%%%%%%%%%%%%%%%%%%%%%%%%%%%%%%%%%%
%%%%%%%%%%%%%%%%%%%%%%%%%%%%%%%%%%%%%%%%%%%%%%%
%%%%%%%%%%%%%%%%%%%%%%%%%%%%%%%%%%%%%%%%%%%%%%%
%%%%%%%%%%%%%%%%%%%%%%%%%%%%%%%%%%%%%%%%%%%%%%%

\begin{flushright} \small
 ITP--UU--10/27 \\ SPIN--10/23 \\ NSF-KITP--10/113
\end{flushright}
\vskip2cm
\begin{center}
 {\large\bfseries New potentials from Scherk-Schwarz reductions}\\[5mm]
Hugo Looyestijn\textsuperscript{1}, Erik Plauschinn\textsuperscript{1,2}, Stefan Vandoren\textsuperscript{1} \\
\bigskip
 {\small\slshape
 \textsuperscript{1} Institute for Theoretical Physics \emph{and} Spinoza Institute \\
 Utrecht University, 3508 TD Utrecht, The Netherlands \\
 \medskip
 \textsuperscript{2} Kavli Institute for Theoretical Physics, Kohn Hall \\
  UCSB, Santa Barbara, CA 93106, USA \\
 \bigskip
 {\upshape\ttfamily H.T.Looijestijn, E.Plauschinn, S.J.G.Vandoren@uu.nl}\\[3mm]}
\end{center}
\vspace{15mm} \hrule\vskip15pt \centerline{\bfseries Abstract}
\smallskip
We study compactifications of eleven-dimensional supergravity on Calabi-Yau threefolds times a circle, with a duality twist along the circle a la Scherk-Schwarz. This leads to four-dimensional $\mathcal N=2$ gauged supergravity with a semi-positive definite potential for the scalar fields, which we derive explicitly. Furthermore, inspired by the orientifold projection in string theory, we define a truncation to $\mathcal N=1$ supergravity. We determine the D-terms, K\"ahler- and superpotentials  for these models and study the properties of the vacua. Finally, we point out  a relation to M-theory compactifications on seven-dimensional manifolds with $G_2$ structure.

\vskip15pt
\hrule\bigskip

\clearpage

%%%%%%%%%%%%%%%%%%%%%%%%%%%%%%%%%%%%%%%%%%%%%%%
%%%%%%%%%%%%%%%%%%%%%%%%%%%%%%%%%%%%%%%%%%%%%%%
%%%%%%%%%%%%%%%%%%%%%%%%%%%%%%%%%%%%%%%%%%%%%%%
%%%%%%%%%%%%%%%%%%%%%%%%%%%%%%%%%%%%%%%%%%%%%%%

\tableofcontents
\addtocontents{toc}{\parskip0ex plus 0.5ex}
\clearpage

%%%%%%%%%%%%%%%%%%%%%%%%%%%%%%%%%%%%%%%%%%%%%%%
%%%%%%%%%%%%%%%%%%%%%%%%%%%%%%%%%%%%%%%%%%%%%%%
%%%%%%%%%%%%%%%%%%%%%%%%%%%%%%%%%%%%%%%%%%%%%%%
%%%%%%%%%%%%%%%%%%%%%%%%%%%%%%%%%%%%%%%%%%%%%%%
%%%%%%%%%%%%%%%%%%%%%%%%%%%%%%%%%%%%%%%%%%%%%%%
%%%%%%%%%%%%%%%%%%%%%%%%%%%%%%%%%%%%%%%%%%%%%%%
%%%%%%%%%%%%%%%%%%%%%%%%%%%%%%%%%%%%%%%%%%%%%%%
%%%%%%%%%%%%%%%%%%%%%%%%%%%%%%%%%%%%%%%%%%%%%%%
%%%%%%%%%%%%%%%%%%%%%%%%%%%%%%%%%%%%%%%%%%%%%%%
%%%%%%%%%%%%%%%%%%%%%%%%%%%%%%%%%%%%%%%%%%%%%%%

\section{Introduction and motivation}

Scherk-Schwarz reductions \cite{Scherk:1978ta,Scherk:1979zr} provide a way to construct gauged supergravities from higher dimensional ungauged ones. They typically lead to semi-positive definite potentials for the scalar fields with local minima that can describe Minkowski or de Sitter vacua. Such models have been studied intensely over recent times in the context of compactifications of string- and M-theory, with and without fluxes. For some background material and earlier references, see e.g. \cite{Bergshoeff:1997mg,Lavrinenko:1997qa,Hull:2002wg,Andrianopoli:2002mf,Dabholkar:2002sy,deWit:2002vt,Andrianopoli:2004im,Dall'Agata:2005ff,Andrianopoli:2005jv,Hull:2005hk,Dall'Agata:2005fm,Hull:2006tp}.

Two classes of Scherk-Schwarz reductions are usually considered: the case of twisted tori (or twistings of the cohomology of other manifolds), and the case of reductions over a circle with a duality twist along the circle. Sometimes, these two classes are related to each other, and reductions with duality twists can be understood in terms of compactifications on twisted tori. For a discussion on this, see e.g. \cite{Hull:2005hk}. This relation will also appear in our investigation, as we will discuss, although we focus primarily on reductions with a duality twist.

In this paper, we present a detailed study of a Scherk-Schwarz
reduction of eleven-dimensional supergravity compactified on a
Calabi-Yau  threefold, denoted by $\cal X$, times a circle, with a
duality twist along the circle. Equivalently, this model can be
formulated as a compactification on a seven-dimensional manifold,
which is a Calabi-Yau fibration over a circle. This yields gauged
$\mathcal N=2$ supergravity in four dimension with a scalar potential for the vector- and hypermultiplet scalars. 
Moreover, there appear Chern-Simons like terms in four dimensions consistent with ${\cal N}=2$ supersymmetry, induced from the Chern-Simons terms in five dimensions.
These models have also been investigated in \cite{Andrianopoli:2004im,Aharony:2008rx}, which we reproduce and elaborate on, and extend to include also the hypermultiplet sector.

The second part of the paper deals with truncations of our models from
$\mathcal N=2$ to $\mathcal N=1$ supersymmetry. 
Inspired by the rules of  the orientifold projection in string theory,
we define a truncation of eleven-dimensional supergravity on ${\cal
  Y}={\cal X} \times S^1$ to $\mathcal N=1$ supergravity in four dimensions. In the absence of the duality twist, our rules are consistent with the results from compactifications of type IIA strings  on Calabi-Yau orientifolds \cite{Grimm:2004ua}. Here, we study  the extension of this truncation to the case when the duality twist is non-trivial. On top of the K\"ahler potential, this yields a  class of superpotentials and D-terms which we compute explicitly. It leads to formulas \eqref{superpot} and \eqref{pot_4_n1_01}, which form one of the 
main new results in this paper. Alternatively, in the picture of the
compactification on the  seven-dimensional manifold $\cal Y$, the
$\mathcal N=1$ supergravity is described by the K\"ahler potential
\begin{equation}
\K = - \log \,\bigl[ \,8    R^3 \,\bigr] 
  -2 \log \left[ 2\hspace{1pt}\V^{\frac 13} R^{-1} \int_{\mathcal Y} \Re \bigl(C \Omega\bigr) \wedge
  \star_7 \Re \bigl(C \Omega\bigr) \right] \;,
\end{equation}
and superpotential
\begin{equation}\label{superpot_introduction}
W= \frac{1}{4}\int_{\cal Y}\,\Big(C_3 +i\sqrt{8}\,\Re (C \Omega) \Big)\wedge {\rm{d}} \Big(C_3 +i\sqrt{8}\,\Re (C \Omega)\Big)\ ,
\end{equation}
where $R$ denotes the radius of the circle, $\V$ is the volume of
the Calabi-Yau threefold $\X$ while $\Omega$ represents its holomorphic three-form, and $C_3$ is the three-form of eleven-dimensional supergravity.
Due to the truncation, $\Omega$ loses some
degrees of freedom and the remaining ones are contained in $\Re
(C\Omega)$, where the compensator $C$ will be defined in~\eqref{eq:def_compensator}. Interestingly, similar formulas for the superpotential have also been obtained in the context of (flux) compactifications of M-theory on $G_2$-manifolds, see e.g. \cite{Beasley:2002db,House:2004pm,Dall'Agata:2005fm}, building on earlier work \cite{Gukov:1999gr,Acharya:2000ps}. This suggests a connection between those models and the ones considered here, which we will discuss in more detail towards the end of this paper.

%%%%%%%%%%%%%%%%%%%%%%%%%%%%%%%%%%%%%%%%%%%%%%%
%%%%%%%%%%%%%%%%%%%%%%%%%%%%%%%%%%%%%%%%%%%%%%%
%%%%%%%%%%%%%%%%%%%%%%%%%%%%%%%%%%%%%%%%%%%%%%%
%%%%%%%%%%%%%%%%%%%%%%%%%%%%%%%%%%%%%%%%%%%%%%%
%%%%%%%%%%%%%%%%%%%%%%%%%%%%%%%%%%%%%%%%%%%%%%%
%%%%%%%%%%%%%%%%%%%%%%%%%%%%%%%%%%%%%%%%%%%%%%%
%%%%%%%%%%%%%%%%%%%%%%%%%%%%%%%%%%%%%%%%%%%%%%%
%%%%%%%%%%%%%%%%%%%%%%%%%%%%%%%%%%%%%%%%%%%%%%%

\section{M-theory on Calabi-Yau manifolds}
\label{sec::cy_geometry}

In this section, we review aspects of compactifications of eleven-dimensional supergravity on Calabi-Yau threefolds.
Almost all material in this section is known, and collected from various places in the literature, which we refer to below. We give this review to recall some of the duality symmetries in five dimensions, and to set our notation for subsequent sections. The reader who is very familiar with five-dimensional matter coupled to ${\cal N}=2$ supergravity might skip this section and go straight to  section \ref{sec:scherk_schwarz} where we present the Scherk-Schwarz reduction to four dimensions.

The low-energy limit of M-theory can be described in terms of eleven-dimensional supergravity. In form-notation, the bosonic part of this action reads~\cite{Cremmer:1978km}
\eq{
  \label{action_m_01}
  \hat{S}
  = \frac{1}{2} \int \left( \hat R \star 1 - \frac{1}{2} \, \hat F_4 \wedge \star \hat F_4 - \frac{1}{6}\,
  \hat F_4\wedge \hat F_4 \wedge \hat C_3 \right) \;.
}
Here,  $\hat R$ denotes the eleven-dimensional Ricci scalar 
and $\star$
stands for the eleven-dimensional Hodge star operator. Furthermore, $\hat
C_3$ is a three-form potential, $\hat F_4=\d \hat C_3$ denotes the
corresponding field strength and we have set the eleven-dimensional Planck constant to one.

In the following, we  compactify M-theory on a simply-connected Calabi-Yau three-fold $\X$, which 
leads to a supergravity theory in five dimensions with eight supercharges~\cite{Cadavid:1995bk}.

%%%%%%%%%%%%%%%%%%%%%%%%%%%%%%%%%%%%%%%%%%%%%%%
%%%%%%%%%%%%%%%%%%%%%%%%%%%%%%%%%%%%%%%%%%%%%%%

\subsection{Calabi-Yau manifolds and dimensional reduction}

\subsubsection*{Notation}

We begin by establishing some notation for the Calabi-Yau three-fold $\mathcal{X}$.
Let us denote a basis of harmonic  $(1,1)$-forms on $\mathcal{X}$ by
\eq{
  \label{basis_1-1}
   \omega_A\;, \hspace{60pt} A=1,\ldots,h^{1,1} \;,
}
where here and in the following $h^{p,q}$ denote the Hodge numbers of the Calabi-Yau threefold.
The triple intersection numbers for $\mathcal{X}$ are defined by
\eq{
  \label{def_tin}
  \mathcal{K}_{ABC} = \int_{\mathcal{X}} \omega_A\wedge\omega_B\wedge
  \omega_C \;.
}
For the third cohomology group $H^3(\mathcal{X})$ we denote a real basis by
\eq{
  \label{basis_3_real}
  \bigl\{\alpha_K ,\beta^L \bigr\}\;,
  \hspace{60pt} K,L=0,\ldots h^{2,1} \;,
}
which is chosen such that 
\eq{
  \label{sym_struct_01}
  \int_{\X}\alpha_K\wedge \beta^L = \delta_K{}^L \;,\hspace{45pt}
  \int_{\X}\alpha_K\wedge\alpha_L = 0\;,\hspace{45pt}
  \int_{\X}\beta^K\wedge \beta^L  = 0 \;.
}

The Calabi-Yau threefold is endowed with a K\"ahler form $J$ and a holomorphic three-form $\Omega$. In terms of the bases \eqref{basis_1-1} and \eqref{basis_3_real}, these can be decomposed in the following way
\eq{
  \label{def_j_o}
  J = v^A \omega_A \;, \hspace{70pt} 
  \Omega  = Z^K \alpha_K - G_K \beta^K \;,
}
where the expansion coefficients $v^A$ are real.
The functions $(Z^K,G_K)$ are the holomorphic sections of
special geometry and depend on the complex structure moduli $z^r$ of the Calabi-Yau manifold, where $r=1,\ldots,h^{2,1}$. 
The volume of $\X$ can be expressed in terms of the K\"ahler form $J$ as follows
\eq{
  \label{def_vol}
  \V = \frac{1}{3!} \int_{\X} J\wedge J\wedge J = \frac{1}{3!} \: \K_{ABC} v^A v^B v^C \;.
}

%%%%%%%%%%%%%%%%%%%%%%%%%%%%%%%%%%%%%%%%%%%%%%%
%%%%%%%%%%%%%%%%%%%%%%%%%%%%%%%%%%%%%%%%%%%%%%%

\subsubsection*{Ansatz for the compactification}
To perform the dimensional reduction of the action~\eqref{action_m_01}, we make the following ansatz for the eleven-dimensional metric
\eq{
  \hat G_{MN} = \left( \begin{array}{cc} \tilde g_{\tilde\mu\tilde\nu} & 0 \\
  0 & G_{mn} \end{array} \right) \;,
  \hspace{60pt}
  \arraycolsep2pt
  \begin{array}{lcl} \tilde\mu,\tilde\nu &=& 0,\ldots, 4 \;, \\[1mm]
  m,n &=& 1,\ldots, 6 \;,
  \end{array}
}
where $\tilde g_{\tilde\mu\tilde\nu}$ denotes a five-dimensional  metric and $G_{mn}$ is the metric of a Calabi-Yau threefold. For the three-form potential, we chose the expansion
\eq{
  \label{ansatz_c3}
  \hat C_3 = \tilde c_3 + A^A \wedge \omega_A + C_3\;,
  \hspace{50pt}
  C_3 =  \sqrt 2\, \xi^K \alpha_K
  - \sqrt 2\, \tilde \xi_K \beta^K \;,
}
with $\tilde c_3(\tilde x^{\tilde\mu})$ a three-form in five
dimensions which depends solely on the five-dimensional coordinates
$\tilde x^{\tilde\mu}$. Similarly, $A^A(\tilde x^{\tilde\mu})$ are
five-dimensional one-forms while $\xi^K(\tilde x^{\tilde\mu})$ and
$\tilde\xi_K(\tilde x^{\tilde\mu})$ are five-dimensional scalars. 
Note that since the  pure Calabi-Yau part $C_3$
features in the superpotential~\eqref{superpot_introduction}, we have separated these terms
from $\tilde c_3$ and $A^A$.

%%%%%%%%%%%%%%%%%%%%%%%%%%%%%%%%%%%%%%%%%%%%%%%
%%%%%%%%%%%%%%%%%%%%%%%%%%%%%%%%%%%%%%%%%%%%%%%

\subsubsection*{Five-dimensional supergravity}
Performing the dimensional reduction to five dimensions  is 
straight-forward and is briefly reviewed  in appendix \ref{app_dim_red}. The resulting five-dimensional low-energy-effective action has been presented in equation \eqref{app_action_5} which we recall for convenience \cite{Gunaydin:1984ak,Bergshoeff:2004kh}
\begin{align}
  \label{action_5}
  \nonumber
  \mathcal{S}_{(5)} = \int_{\mathbb{R}^{4,1}} \biggl[ 
  \; &+\frac{1}{2}\,R_{(5)} \star_5 1 - \frac{1}{4}\, \d \log\V \wedge\star_5 \d\log\V
   + \frac 14 \K_{ABC} \nu^C  \d\nu^A \wedge\star_5 \d \nu^B \\
  \nonumber
   &    + \frac 14 \Bigl(\K_{ABC}\nu^C -\frac 14 \K_{ACD} \nu^C \nu^D
   \K_{BEF} \nu^E \nu^F\Bigr) \, \d A^A\wedge\star_5 \d A^B 
    \\
    & -\frac{1}{12}\: \K_{ABC}\,\d A^A\wedge \d A^B\wedge A^C 
    - G_{r\ov s} \d z^r \wedge\star_5 \d \ov z^{\ov s}  \\
  \nonumber
    &  -\frac{1}{4\V^2} \, \Bigl( \d a +  \xi^K\d\tilde\xi_K  - \tilde\xi_K\d\xi^K  
   \Bigr) \wedge \star_5 \Bigl( \d a + \xi^L\d\tilde\xi_L - \tilde\xi_L\d\xi^L 
    \Bigr) \\
  \nonumber    
    & +\frac{1}{2\V} \bigl( {\rm Im}\,\mathcal{M} \bigr)^{-1\, KL}
    \Bigl( \d\tilde\xi_K - \mathcal{M}_{KN}  \d\xi^N \Bigr)  \wedge\star_5
    \Bigl( \d\tilde\xi_L - \ov{\mathcal{M}}_{LM}  \d\xi^M \Bigr)   
    \;\; \biggr] \;.
\end{align}
The first term in this expression is the five-dimensional Ricci
scalar, $\V$ is the volume of the Calabi-Yau manifold and $\K_{ABC}$
denote the triple intersection numbers defined in \eqref{def_tin}. The
 matrix $G_{r\ov s}$ as well as the period matrix $\mathcal{M}$ have been introduced
in appendix \ref{app_dim_red}. 

The scalars $\nu^A$ are related to the expansion coefficients
$v^A$ of the K\"ahler form $J$ by 
a rescaling with the volume (see equation \eqref{app_nus}),
such that they 
satisfy 
\eq{
  \label{restriction_nu}  
\frac 16 \, \K_{ABC} \nu^A \nu^B \nu^C =1\;.
}
Thus, there are  $h^{1,1}-1$ scalar degrees of freedom in
these fields. 
Accordingly,  the vector fields $A^A$ comprise the
graviphoton and $h^{1,1}-1$ additional vector fields to form five-dimensional vector multiplets. The remaining scalar fields $\{\V, a, z^r , \ov z^r, \xi^K, \tilde
\xi_K\}$ form $h^{2,1}+1$ hypermultiplets that parametrize a quaternion-K\"ahler manifold \cite{Cadavid:1995bk}.

%%%%%%%%%%%%%%%%%%%%%%%%%%%%%%%%%%%%%%%%%%%%%%%
%%%%%%%%%%%%%%%%%%%%%%%%%%%%%%%%%%%%%%%%%%%%%%%
%%%%%%%%%%%%%%%%%%%%%%%%%%%%%%%%%%%%%%%%%%%%%%%
%%%%%%%%%%%%%%%%%%%%%%%%%%%%%%%%%%%%%%%%%%%%%%%

\subsection{Symmetries of the five-dimensional theory}
\label{sect:symmetries}

%%%%%%%%%%%%%%%%%%%%%%%%%%%%%%%%%%%%%%%%%%%%%%%
%%%%%%%%%%%%%%%%%%%%%%%%%%%%%%%%%%%%%%%%%%%%%%%

\subsubsection{Symmetries in the vector multiplet sector}

We begin our discussion on the symmetries of~\eqref{action_5} with the vector multiplets. Besides the usual gauge invariances acting on the vector potentials, there are additional symmetries in the scalar sector. In particular, the scalars in the vector multiplets parametrize a so-called real special geometry, whose isometries have been studied in \cite{Gunaydin:1984ak}. As explained in \cite{deWit:1992cr}, not all isometries extend to symmetries of the full Lagrangian, but only  transformations 
\eq{
  \label{symmetry_vector_5}
  \delta \nu^A = M^A{}_B  \nu^B 
  \;,\hspace{50pt}
  \delta A^A = M^A{}_B A^B \;,
}
where the constant, real matrix $M^A{}_B$ is subject to the constraint
\eq{\label{eq:cycl_condition}
  0 =  \K_{D(AB}M^D{}_{C)} = \K_{DBC} M^D{}_A  +  \K_{ADC} M^D{}_B +  \K_{ABD}M^D{}_C \;,
}
lead to symmetries of the full action, 
including the Chern-Simons terms.

Generically, the real special manifolds parametrized by the scalars in the vector multiplets need not be homogeneous, and solutions to \eqref{eq:cycl_condition} are not known in general. However, for homogeneous spaces a classification  can be found in 
\cite{deWit:1991nm,deWit:1992wf}. A special subclass of the latter is given by the manifolds
\begin{equation}
SO(1,1)\times \frac{SO(n+1,1)}{SO(n+1)}\ ,
\end{equation}
for any integer $n$,
with isometry group $SU(1,1)\times SO(n+1,1)$. This case arises in compactifications in which the Calabi-Yau manifold is  a $K3$-fibration over a base $P^1$. In the present context, this situation has been studied in \cite{Aharony:2008rx}.

%%%%%%%%%%%%%%%%%%%%%%%%%%%%%%%%%%%%%%%%%%%%%%%
%%%%%%%%%%%%%%%%%%%%%%%%%%%%%%%%%%%%%%%%%%%%%%%

\subsubsection{Symmetries in the hypermultiplet sector}

\paragraph*{Notation}
To study the isometries for the hypermultiplets, we first
introduce some notation. The hypermultiplet scalars were given by  $\{\V, a, z^r, \ov z^r, \xi^K, \tilde
\xi_K\}$, which  parametrize a particular type of quaternionic
manifolds called `very special' in \cite{deWit:1992wf}. 

Since we consider M-theory on a Calabi-Yau manifold,  the subspace of
complex structure deformations
$z^r$ is described by special K\"ahler geometry, for which there exists a
prepotential.
In the large complex structure limit, it is given by\,\footnote{We reserve the
  usual notation $F$ and $X$ for the special geometry in the vector multiplets.}
\eq{
  \label{prepot_01}
  G(Z) = - \frac{1}{3!}\: d_{rst} \frac{Z^r Z^s Z^t}{Z^0} \;,\hspace{55pt}
  r,s,t=1,\ldots,h^{2,1} \;.
}
Here, $d_{rst}$ is a real symmetric tensor,  the  $Z^K$ appear in
the expansion \eqref{def_j_o} of the holomorphic three-form
$\Omega$. The connection to the scalars $z^r$ is made by introducing projective coordinates 
\eq{
  \label{def_z}
  z^r = \frac{Z^r}{Z^0} \;,\hspace{70pt} r=1,\ldots, h^{2,1} \;.
}
The corresponding K\"ahler potential reads
\eq{
  \label{kaehler_pot_01}
  {\mathcal K^{\rm cs}} = - \ln \left( i \int_{\X} \Omega\wedge\ov \Omega \right) 
  = - \ln \left( \,\frac{4}{3} \, \bigl\lvert Z^0 \bigr\rvert^2 \:d \right) \;,
}
where here and in the following we employ the notation
\eq{
  \label{notation_tin_01}
  d = d_{rst} x^rx^sx^t \;,\hspace{40pt}
  d_r = d_{rst} x^sx^t \;,\hspace{40pt}
  d_{rs} = d_{rst} x^t \;,
}
with $x^r = {\rm Im}\, z^r$. From \eqref{kaehler_pot_01}, we can then compute the K\"ahler metric as\,\footnote{The identification of \eqref{metric_01} with the metric \eqref{app_metric_cs} can be made by noting that $\chi_r =  \partial_{z^r}\Omega + \bigl( \partial_{z^r}  \mathcal{K}^{\rm cs}  \bigr) \Omega$ as well as that $\int_{\X} \partial_{z^r}\Omega\wedge\ov\Omega =0$.}
\eq{
  \label{metric_01}
  G_{r\ov s} = \frac{\partial^2}{\partial z^r \partial \ov z^{\ov s}} \:
  \mathcal{K^{\rm cs}} = -\frac{3}{2} \: \frac{d_{rs}}{d}
  + \frac{9}{4} \frac{d_r d_s}{d^2} \;.
}
With $G^{r\ov s}$ denoting the inverse of \eqref{metric_01}, the
curvature for this metric can be computed as follows
\cite{deWit:1992wf}
\eq{
  R^r{}_{st}{}^v = \delta^r_s \delta^v_t + \delta^r_t\delta^v_s - \frac{4}{3}\, C^{rvu} d_{stu} \;,
  \qquad{\rm where}\qquad \!\!\! C^{rst} = \frac{27}{64} \,\frac{1}{d^2}\,
  G^{r\ov u} G^{s\ov v} G^{t\ov w} \, d_{uvw} \,.
}

%%%%%%%%%%%%%%%%%%%%%%%%%%%%%%%%%%%%%%%%%%%%%%%
%%%%%%%%%%%%%%%%%%%%%%%%%%%%%%%%%%%%%%%%%%%%%%%

\paragraph*{Symmetries for $\mathbf{z^r}$}

Since the scalars $z^r$ appearing in the action \eqref{action_5} can
be described by a K\"ahler potential, their kinetic term  is invariant
provided that \eqref{kaehler_pot_01} does not change under the
transformations of interest.\footnote{Strictly speaking,
\eqref{kaehler_pot_01} should be invariant up to K\"ahler
transformations, but we will ignore those in the present analysis.}  We then make the following ansatz for the transformation of the
sections $(Z^K,G_K)$ appearing in the holomorphic three-form $\Omega$
\eq{
  \label{twist_3_01}
  \delta \binom{Z^K}{G_K} =  
  \left( \begin{array}{cc} \Q^K{}_L & \R^{KL} \\ \s_{KL} & \T_K{}^L \end{array}\right)
  \binom{Z^L}{G_L}  \;,
}
where, $\Q$, $\R$, $\s$ and $\T$ are constant, real, 
square matrices of dimension $h^{2,1}+1$. Imposing the invariance of the K\"ahler potential~\eqref{kaehler_pot_01} under this transformation, i.e.
\eq{
  \delta \int_{\X} \Omega\wedge\ov\Omega = 0 \;,
}
we are lead to the constraints
\eq{
  \T = - \Q^T \;,\hspace{60pt} \s=\s^T \;,\hspace{60pt} \R=\R^T \;,
}
which means that these isometries have to be contained in the symplectic group
$Sp\bigl( 2(h^{2,1} + 1), \mathbb R\bigr)$.
However, because we are considering a Calabi-Yau manifold, we know that
the sections $G_K$ are related to  $Z^K$ through  a prepotential $G(Z)$ as $G_K=\partial G(Z)/\partial Z^K$. Therefore, in the ansatz \eqref{twist_3_01} the transformation $\delta G_K$ is not independent of $\delta Z^K$, but we have to require
\eq{
  \label{cons_10}
  \delta G_K = \frac{\partial G_K}{\partial Z^L} \,\delta Z^L \;.
}
Recalling that $G_K$ is a homogeneous function of degree one in
the $Z^K$, that is  $(\partial G_K/\partial Z^L)Z^L = G_K$, we
infer from \eqref{cons_10} that \cite{deWit:1992wf}
\eq{
  \label{cons_11}
   0 = G^T \mathcal{Q} \,Z + G^T \mathcal{R} \,G - Z^T \mathcal{S}\, Z - Z^T \mathcal{T} \,G \;,
}
where matrix multiplication is understood.
Furthermore, to leading order in the large $z^r$-expansion, for Calabi-Yau
threefolds the prepotential $G(Z)$ is given by~\eqref{prepot_01}.  The solution to
\eqref{cons_11} in this case can be found in  \cite{deWit:1992wf} which
we briefly recall. In particular, the matrices $\cal Q,R,S$ and
$\mathcal T$ appearing in \eqref{twist_3_01} can be parametrized as
\eq{
  \label{twist_cubic}
  & \mathcal{Q}^K{}_L = - \bigl(\mathcal{T}^T\bigr)^K{}_L  =  \left( \begin{array}{cc}
    \beta & a_s \\ b^r & B^r{}_s + \frac{1}{3}\, \beta \,\delta^r{}_s \end{array}\right) \;, \\[3mm]
  & \mathcal{S}_{KL}  = -\left( \begin{array}{cc}
    0 & 0 \\ 0 & d_{rst} b^t \end{array}\right) \;,\hspace{40pt} 
  \mathcal{R}^{KL}  = - \left( \begin{array}{cc}
    0 & 0 \\ 0 & \frac{4}{3}\, C^{rst} a_t \end{array}\right) \;,
}
with $\beta$, $b^r$, $a_s$ and $B^r{}_s$ constant parameters. The matrix $B^r{}_s$ is subject to the constraint
\eq{
  B^r{}_{(s} d_{tu)v} = 0 \;,
}
where $(\cdot \cdot \cdot)$ denotes symmetrization and the constants $a_s$ are constrained by
\eq{
\label{condition_with_E}
  0 = a_s\, E^s_{tuvw} \hspace{35pt}{\rm where}\hspace{35pt}
  E^s_{tuvw} = C^{yzs}\, d_{y(tu} d_{vw)z} - \delta^s\!{}_{(t} d{}_{uvw)} \;.
}
With this information, we can compute the transformation of the projective coordinates $z^r$ introduced in \eqref{def_z}. Employing \eqref{twist_cubic} as well as \eqref{def_z}, we find  \cite{deWit:1992wf}
\eq{
  \label{ztwist}
  \delta z^r = b^r - \frac{2}{3} \,\beta\, z^r + B^r{}_s z^s -\frac{1}{2}\, R^r{}_{st}{}^v 
  z^s z^t a_v \;,
}
and we note that  the condition~\eqref{condition_with_E} implies that $R^r{}_{st}{}^v a_v$
is constant.

%%%%%%%%%%%%%%%%%%%%%%%%%%%%%%%%%%%%%%%%%%%%%%%
%%%%%%%%%%%%%%%%%%%%%%%%%%%%%%%%%%%%%%%%%%%%%%%

\paragraph*{Symmetries for $\mathbf{\xi^K}$ and $\mathbf{{\tilde \xi}_K}$}

To promote the symmetry of the complex structure deformations $z^r$ to
a symmetry for the full hypermultiplets, and hence to isometries of the quaternionic space, 
we follow again \cite{deWit:1992wf}. First we 
note that the period matrix $\mathcal{M}$ appearing in the action
\eqref{action_5} (as well as in equations \eqref{app_intersections})
satisfies the relation
\eq{
  G_K = \mathcal{M}_{KL} Z^L \;.
}
From the transformation of $(Z^K,G_K)$ shown in \eqref{twist_3_01}, we infer that $\mathcal{M}$ transforms as
\eq{
  \delta \mathcal{M} = \s + \T \mathcal{M} - \mathcal{M} \Q - \mathcal{M} \R \mathcal{M} \;.
}
Requiring  the kinetic term of the scalars $(\xi^K,\tilde\xi_K)$ in \eqref{action_5} to be invariant implies their following transformation 
\eq{
\label{eq:transf_xi}
  \delta \binom{\xi^K}{\tilde\xi_K} =
  \left( \begin{array}{cc} \Q^K{}_L & \R^{KL} \\ \s_{KL} & \T_K{}^L \end{array}\right)
  \binom{\xi^L}{\tilde\xi_L} \;,
}
which also leads to the invariance of the
$\bigl(\xi^K\d\tilde\xi_K-\tilde\xi_K\d\xi^K\bigr)$ terms and agrees
with~\cite{deWit:1992wf}. Hence, just like $(Z^K,G_K)$, the $(\xi^K,
\tilde \xi_K)$ form a symplectic pair.

Finally, we should add that the hypermultiplet space in general possesses more
symmetries than the ones described here, for instance the Heisenberg algebra of isometries (which include the Peccei-Quinn shifts on $(\xi^K,{\tilde \xi}_K)$) that
act on the coordinates $(\xi^K,{\tilde \xi}_K)$ and $a$ only. Furthermore, there are additional isometries that 
act non-trivially on the volume  $\mathcal V$ and the axion $a$ -- for a complete classification see \cite{deWit:1992wf}. 
Including these in a Scherk-Schwarz reduction would be an interesting extension of our work. We will not consider them in our present discussion.

%%%%%%%%%%%%%%%%%%%%%%%%%%%%%%%%%%%%%%%%%%%%%%%
%%%%%%%%%%%%%%%%%%%%%%%%%%%%%%%%%%%%%%%%%%%%%%%
%%%%%%%%%%%%%%%%%%%%%%%%%%%%%%%%%%%%%%%%%%%%%%%
%%%%%%%%%%%%%%%%%%%%%%%%%%%%%%%%%%%%%%%%%%%%%%%

\section{Scherk-Schwarz reduction to four dimensions}
\label{sec:scherk_schwarz}

In this section, we  compactify the
five-dimensional theory given by~\eqref{action_5} on a circle of radius $R$. In addition,
we  impose a non-trivial dependence on the coordinate of the
circle. Such a setup was studied first in~\cite{Andrianopoli:2004im}
and, without hypermultiplets, further worked out in~\cite{Aharony:2008rx}.

%%%%%%%%%%%%%%%%%%%%%%%%%%%%%%%%%%%%%%%%%%%%%%%
%%%%%%%%%%%%%%%%%%%%%%%%%%%%%%%%%%%%%%%%%%%%%%%
%%%%%%%%%%%%%%%%%%%%%%%%%%%%%%%%%%%%%%%%%%%%%%%
%%%%%%%%%%%%%%%%%%%%%%%%%%%%%%%%%%%%%%%%%%%%%%%

\subsection{Ansatz for the compactification}

To perform the compactification from five to four dimensions, we split the five-dimensional coordinates as
\eq{
  \{\tilde x^{\tilde \mu} \}\longrightarrow \{ x^{\mu} , z \} \;,\hspace{50pt}
  \arraycolsep2pt
  \begin{array}{lcl}
  \tilde\mu &=& 0,\ldots, 4 \;, \\[2pt]
  \mu &=& 0,\ldots, 3 \;,
  \end{array}
}
where $z$ denotes the coordinate of the circle normalized as $z\sim z+1$.
The dependence of the five-dimensional scalars $\nu^A$ and the five-dimensional vectors $A^A$ on the coordinate $z$ is chosen in the following way
\begin{align}
  \label{def_twist_vec}
  \partial_z \nu^A = M^A{}_B \nu^B\,, \hspace{60pt}
  \partial_z A^A = M^A{}_B A^B\,,
\end{align}
where $M^A{}_B$ satisfies~\eqref{eq:cycl_condition}. These expressions can be integrated
to obtain
\begin{align}\label{def_twist_vec_finite}
  \nu^A(z) = \Bigl[\exp(Mz)\Bigr]^A_{\,\,\,\,B} \, \nu^B (0)\;, \hspace{40pt}
  A^A(z) = \Bigl[\exp(Mz)\Bigr]^A_{\,\,\,\,B} \, A^B(0)\;,
\end{align}
where the exponential of the matrix $M$ is understood as a matrix product and where only the $z$-dependence of the fields is shown explicitly.

Clearly, the fields are not periodic around the circle, but are related to each other by the duality transformations \eqref{symmetry_vector_5} generated by  $M$. These duality transformations form a group $G$, and therefore one should have
\begin{equation}
\exp(M) \in G\ .
\end{equation}
Classically, the group $G$ is taken over the real numbers, and hence
the entries of $M$ can be taken as arbitrary real constants. They determine the masses of the fields in four dimensions, and are treated as continuous parameters which we can take to be arbitrary small  -- or at least to be smaller than the masses of the Kaluza-Klein (KK) modes that we neglected. In the quantized theory, however, we expect the duality group to be defined over the integers, and hence the masses will be quantized in some units. This no longer guarantees that they are smaller than the masses of the KK modes. In turn, this  could lead to complications in the truncation of the theory to the lightest modes, which we will ignore in this paper. For discussions on this issue for toroidal compactifications, see for instance \cite{Dabholkar:2002sy,Hull:2005hk}. Essentially, this problem is similar to what one encounters in flux compactifications, where one has to make sure that there is a separation of mass scales, in particular the mass scale induced by the fluxes and the KK mass scale.

After this important side comment, we now turn to the hypermultiplets. For the dependence of the scalars $(\xi^K,\tilde\xi_K)$ on the coordinate $z$ of the circle we take
\begin{align}
   \label{twist_3_05}
   \partial_z \begin{pmatrix} \xi^K  \\ \tilde \xi_K \end{pmatrix}
&=
\begin{pmatrix} \mathcal Q^K{}_L & \mathcal R^{KL} \\ \mathcal S_{KL}
  & \mathcal T_K{}^L \end{pmatrix}
\begin{pmatrix} \xi^L  \\ \tilde \xi_L\end{pmatrix} \;,
\end{align}
and for the complex structure moduli $z^r$ we choose in a similar fashion
\eq{
  \label{def_curl_n}
  \partial_z z^r &= b^r - \frac{2}{3} \,\beta\, z^r + B^r{}_s z^s -\frac{1}{2}\, R^r{}_{st}{}^v a_v 
  z^s z^t
  \equiv \mathcal{N}^r \,.
}
The finite version of these transformations can easily be written down
for $(\xi^K, \tilde \xi_K)$. For $z^r$, one first expresses them
as transformations for the sections $Z^K$, after which one can integrate. For the scalars $a$ and $\V$, we choose
\eq{
  \partial_z a = 0 \;,\hspace{60pt}
  \partial_z \V = 0 
  \;. 
}

Note that, since we have chosen the dependence of the fields on  the
circle coordinate $z$ such that they correspond to Killing vectors of the five-dimensional theory, the full action does not depend on $z$ and so we can evaluate the terms at a particular reference point, say $z_0=0$.

For the five-dimensional metric, we make the following ansatz for the dimensional reduction
\eq{
  \label{metric_5_01}
  \tilde g_{\tilde \mu\tilde\nu} = \left( \begin{array}{cc} R^{-1} g_{\mu\nu} + R^2 
  A^0_{\mu}A^0_{\nu} & 
  -R^2 A^0_{\mu} \\  -R^2 A^0_{\nu} & R^2 \end{array} \right) \;,
}
where $g_{\mu\nu}$ is the four-dimensional metric, $R$ is the radius
of the circle and where the four-vector $A^0_{\mu}$ will become the
graviphoton. The factor $R^{-1}$ is chosen such that we end up in
Einstein frame. For the five-dimensional gauge fields  appearing in the action \eqref{action_5}, we choose
\begin{align}
    \label{vec_5_4}
    A^A_{(5)} = A^A_{(4)} + b^A \bigl( {\rm d}z -  A^0 \bigr) \;, 
\end{align}
where we added subscripts to distinguish between five- and four-dimensional quantities.
Using the above ans\"atze within the action \eqref{action_5}, one can perform the dimensional reduction, which is outlined in appendix \ref{app_dim_red_4}. Below, we present the results.

%%%%%%%%%%%%%%%%%%%%%%%%%%%%%%%%%%%%%%%%%%%%%%%
%%%%%%%%%%%%%%%%%%%%%%%%%%%%%%%%%%%%%%%%%%%%%%%
%%%%%%%%%%%%%%%%%%%%%%%%%%%%%%%%%%%%%%%%%%%%%%%
%%%%%%%%%%%%%%%%%%%%%%%%%%%%%%%%%%%%%%%%%%%%%%%

\subsection{The four-dimensional action}
\label{sec:n2sugra}

The dimensional reduction from five to four dimensions as well as  bringing the result into the standard form of $\mathcal N=2$ supergravity  can be found in appendix \ref{app_dim_red_4}. In particular, the four-dimensional action takes the form
\eq{
  \label{eq:n2lagr}
  \mathcal{S}_{(4)} = \int_{\mathbb{R}^{3,1}} \biggl[ \qquad&\frac{1}{2}\: R_{(4)} \star_4 1 
   + \frac{1}{4} \, {\rm Im}\, \mathcal{N}_{\Lambda\Sigma} \, F^{\Lambda} \wedge\star_4 
   F^{\Sigma}
   +\frac{1}{4} \, {\rm Re}\, \mathcal{N}_{\Lambda\Sigma} \, F^{\Lambda} \wedge
   F^{\Sigma} \\
   -\,&g_{A \ov B}\: Dt^A\wedge\star_4 D\ov t^B 
   - \frac{1}{6} \: A^A M_A{}^B \wedge A^C \wedge \d A^D \K_{BCD} \\
   -\,&   h_{uv} D q^u \wedge \star_4 D q^v - V \quad\biggr] \;,   
}
where $\Lambda,\Sigma=0,\ldots, h^{1,1}$ while $A,B,\ldots=1,\ldots, h^{1,1}$, and
where we have omitted labels on the vector fields indicating four-dimensional
quantities. We have furthermore defined
\eq{\label{eq:redefinitions_d4}
\phi^A = R\,  \nu^A \hspace{60pt}{\rm and}\hspace{60pt}
\mathcal \V = {\rm e}^{-2\phi} \;,
}
as well as the complexified K\"ahler moduli and their derivatives
\eq{
  \label{def_t_01}
  t^A = b^A + i \, \phi^A \;,\hspace{60pt}
  Dt^A = \d t^A -  M^A{}_B\bigl(A^B - t^B A^0 \bigr)\;,
}  
where $b^A$ appeared in \eqref{vec_5_4}. The K\"ahler metric $g_{A\ov B}$ is written as
\eq{
  \label{metric_16}
  g_{A\ov B} 
  = - \frac{1}{4R^3} \left( \K_{AB} - \frac{\K_A \K_B}{4R^3}\right) \;,
}
where we have employed the following notation 
\eq{
 \K_A = \K_{ABC} \phi^B \phi^C \;,\hspace{60pt}
    \K_{AB} = \K_{ABC} \phi^C \;,
}
with $\K_{ABC}$ the triple intersection numbers defined in
\eqref{def_tin}. Using these as well as \eqref{eq:redefinitions_d4} in the constraint \eqref{restriction_nu}, we also find
\eq{
\label{relation_r_45}
R^3 = \frac{1}{6}\: \K_{ABC} \phi^A \phi^B \phi^C\;.
}
The metric \eqref{metric_16} is a special K\"ahler metric and can be derived from
the following prepotential 
\begin{align}\label{eq:special-prepotential}
  F = -\frac 1 {3!}\, \K_{ABC}\, \frac{X^A X^B X^C}{X^0}\;, \hspace{50pt}A,B,C=1,\ldots,h^{1,1} \;,
\end{align}
where we employ coordinates $\{X^0,X^A\}$ with $X^A = X^0 \,t^A$. The corresponding  K\"ahler potential reads
\begin{align}\label{eq:special-kahler}
  \K^{\rm vec} \equiv -\log \Bigl[ \,i \,\overline X^\Lambda F_\Lambda - i\,
    X^\Sigma \overline F_\Sigma \, \Bigr] = - \log \,\bigl[ \,8    R^3 \,\bigr]\;,
\end{align}
where due to the symmetries of the theory we can set $X^0=1$. 
The expressions for the period matrix  $\mathcal
N_{\Lambda\Sigma}$ are given in \eqref{def_im_n}, and the  field strengths appearing in \eqref{eq:n2lagr} are written as
\eq{
  \label{def_vec_01}
  F^{\Lambda} = {\rm d} A^{\Lambda} + \frac{1}{2} \, f^{\Lambda}{}_{\Sigma\Gamma}
  A^{\Sigma} \wedge A^{\Gamma} \;,\hspace{60pt} \Lambda,\Sigma,\Gamma = 0, \ldots,h^{1,1} \;.
}  
The structure constants are \cite{Andrianopoli:2004im,Aharony:2008rx}
\eq{
  \label{def_vec_02}
  f^0{}_{AB} = 0 \;,\hspace{40pt}
  f^C{}_{AB} = 0 \;,\hspace{40pt}
  f^B{}_{A0} = - M^B{}_A \;,
}
and they define the gauge group which we elaborate on in the next subsection. 

We mention here that gauge invariance of the action \eqref{eq:n2lagr} requires the presence of Chern-Simons-like terms, which are inherited from the five-dimensional Chern-Simons term. These arise when the matrix ${\rm Re}\, \mathcal{N}_{\Lambda\Sigma}$ transforms nontrivially under the action of the gauge group, in such a way that it needs to be compensated by an additional term in the action, the last term on the second line in \eqref{eq:n2lagr}. The existence of such terms in gauged supergravity was found in \cite{deWit:1984px}, and in the present context it was discussed in \cite{Aharony:2008rx}. Some further applications of these terms in the study of ${\cal N}=2$ supersymmetric vacua can be found in \cite{Hristov:2009uj}.

Turning to the hypermultiplet sector, we find that it is described by
\eq{
  \label{result_hypers_01}
  h_{uv} D_\mu q^u &D^\mu q^v =  G_{r\ov s}\, D_{\mu}z^r D^\mu \ov
  z^{\ov s}
  + \partial_\mu \phi\, \partial^\mu \phi \\[2mm]
  +\,&\frac {{\rm e}^{4\phi}} 4 \Bigl( \partial_{\mu} a +  \xi^K D_{\mu} \tilde\xi_K - \tilde\xi_K  D_{\mu}\xi^K
       \Bigr) \Bigl( \partial^{\mu} a +  \xi^L D^{\mu} \tilde\xi_L - \tilde\xi_L  D^{\mu}\xi^L
       \Bigr) \\
   -\, & \frac{{\rm e}^{2\phi}}{2} \bigl( {\rm Im}\,\mathcal{M}    \bigr)^{-1\, KL}
    \Bigl( D_{\mu}\tilde\xi_K - \mathcal{M}_{KP}  D_{\mu}\xi^P \Bigr) 
    \Bigl( D^{\mu}\tilde\xi_L - \ov{\mathcal{M}}_{LQ}  D^{\mu}\xi^Q \Bigr) \,,
}
where $\mu=0,\ldots,3$ and $G_{r\ov s}$ has been introduced in \eqref{metric_01}. The covariant derivatives appearing here are
\eq{
\label{eq:cov-dev-hypers}
  D_{\mu}z^r = \partial_{\mu} z^r - \mathcal N^r A_{\mu}^0\;, 
  \hspace{40pt}
  D_{\mu} \binom{\xi}{\tilde\xi} =  \partial_{\mu} \binom{\xi}{\tilde\xi}
  - N\,  \binom{\xi}{\tilde\xi} \,A_{\mu}^0 \;,
}
where $\mathcal{N}^r$ had been defined in \eqref{def_curl_n}, where appropriate indices for $(\xi,\tilde\xi)$ are understood  and where the matrix $N$ is given by 
\eq{
  \label{eq:definition-twist-n}
  N = \begin{pmatrix} \mathcal Q^K{}_L & \mathcal R^{KL} \\ \mathcal S_{KL}
  & \mathcal T_K{}^L \end{pmatrix} \;.
}
Finally, the scalar potential can be expressed in the following way 
\eq{
  \label{potential_4}
  V =\; \quad&
   \frac{1}{R^3}\: \mathcal{N}^r \ov{\mathcal{N}}^s
     \, G_{r\ov s} 
  + \frac{\rm {e}^{4\phi}}{4R^3} \left[ \:
    \binom{\xi}{\tilde \xi}^TN^T\:\binom{\tilde\xi}{-\xi} \:   \right]^2 \\
  -\,& \frac{\rm {e}^{2\phi}}{2R^3} \: 
    \binom{\xi}{\tilde\xi}^T  N^T \:\binom{\mathcal{M}}{-\mathds{1}}\:
    \bigl( {\rm Im}\,\mathcal{M}\bigr)^{-1}\,
    \binom{\ov{\mathcal{M}}}{-\mathds{1}}^T
    N\:
    \binom{\xi}{\tilde\xi} \\[2mm]
  -\,& \frac{1}{4 R^6}\: \bigl( M^A{}_C \phi^C \bigr)\,\bigl(M^B{}_D \phi^D\bigr)\: \K_{AB} \;,
}
where matrix multiplication with correct contraction of indices is again understood.
We will study the properties of this potential in section 6.

%%%%%%%%%%%%%%%%%%%%%%%%%%%%%%%%%%%%%%%%%%%%%%%
%%%%%%%%%%%%%%%%%%%%%%%%%%%%%%%%%%%%%%%%%%%%%%%
%%%%%%%%%%%%%%%%%%%%%%%%%%%%%%%%%%%%%%%%%%%%%%%
%%%%%%%%%%%%%%%%%%%%%%%%%%%%%%%%%%%%%%%%%%%%%%%

\subsection{Gauged $\mathcal N=2$ supergravity formulation}
The ungauged part of the Lagrangian~\eqref{eq:n2lagr} is already written in the usual form
of four-dimensional $\mathcal N=2$ supergravity. The only changes we have to
explain are the modifications due to the gauging, in particular the
covariant derivatives for the scalars, and  the scalar potential. 

%%%%%%%%%%%%%%%%%%%%%%%%%%%%%%%%%%%%%%%%%%%%%%%
%%%%%%%%%%%%%%%%%%%%%%%%%%%%%%%%%%%%%%%%%%%%%%%

\subsubsection*{Covariant derivatives, Killing vectors and isometries}

The
covariant derivatives are given by
\eq{
  D_\mu q^u =\partial_\mu q^u + \tilde k^u_\Lambda A^\Lambda_\mu\;, \hspace{60pt}
  D_\mu t^A =\partial_\mu t^A + k^A_\Lambda A^\Lambda_\mu\;,
}
where the quantities $\tilde k^u_\Lambda$
and $k^A_\Lambda$ are Killing vectors on the quaternionic and
special K\"ahler spaces, respectively. For the scalars $t^A$ in the vector multiplets, from~\eqref{def_t_01} we  read off  that
\begin{align*}
  k^A_0 = M^A{}_B t^B\;,\hspace{60pt} k^A_B = -M^A{}_B\,,
\end{align*}
which  means that on the special K\"ahler space defined by \eqref{eq:special-prepotential}, the isometries  we are gauging are given by
\begin{equation}\label{gauge-group}
\delta t^A=- M^A{}_B a^B+
 M^A{}_Bt^B a^0\ ,
\end{equation}
for some arbitrary parameters $a^0$ and $a^A$.
That these are indeed isometries follows from the analysis of the special K\"ahler subsector of the hypermultiplets given in \eqref{ztwist}, which is completely analogous. Here, the symmetries \eqref{gauge-group} 
correspond to the first and third term in \eqref{ztwist}, namely a shift in $t^A$ and a linear transformation with a matrix satisfying \eqref{eq:cycl_condition}. 
The gauge group is thus a subgroup of the duality group of isometries on the special K\"ahler manifold. This duality group contains the one from the five-dimensional theory, but  in four dimensions it gets extended to a larger group \cite{deWit:1992wf}. The structure constants of the gauge group are given by \eqref{def_vec_02}, and define a solvable Lie algebra which is the semi-direct product of two Abelian subalgebras of dimension one (graviphoton) and $h^{1,1}$ (the other vector potentials) \cite{Andrianopoli:2004im,Aharony:2008rx}.

The isometry group for the hypermultiplets can easily be read off from 
\eqref{eq:cov-dev-hypers}. 
It is a $U(1)$ group, realized linearly on the scalars $(\xi^K,\tilde\xi_K)$ but  non-linearly on  $z^r$ (see  equation \eqref{def_curl_n}). 
The gauge group 
acts on it only via the graviphoton. 

%%%%%%%%%%%%%%%%%%%%%%%%%%%%%%%%%%%%%%%%%%%%%%%
%%%%%%%%%%%%%%%%%%%%%%%%%%%%%%%%%%%%%%%%%%%%%%%

\subsubsection*{Scalar potential}

The explicit form of the scalar potential is given in
\eqref{potential_4}. It  can be written in the standard form of
$\mathcal N=2$ supergravity which reads\,\footnote{\label{footnote::factor2}The
 overall factor $2$ compared to the potential
 of~\cite{Andrianopoli:1996cm} is due to the different normalization
 in~\eqref{eq:n2lagr}. When rescaling the four-dimensional metric
 in~\eqref{eq:n2lagr} as $g \to  \frac 12 g$, one arrives at 
 the form of~\cite{Andrianopoli:1996cm}.}
\begin{align}\label{eq:formal-potential}
  V = 2\, {\rm e}^{\mathcal K^{\rm vec}} \Bigl(4\, h_{uv} \tilde k^u_\Lambda \tilde
  k^v_\Sigma +  g_{A\ov B} \,k^A_\Lambda\,
{\overline k}{}^B_\Sigma\,\Bigr)\, \overline X{}^\Lambda X^\Sigma\;,
\end{align}
where $\mathcal K^{\rm vec}$ and $g_{A\ov B}$ were defined
in~\eqref{eq:special-kahler} and~\eqref{metric_16}, respectively. 
In the general expression for the $\mathcal N=2$ scalar potential, there is an additional term proportional to 
the quaternionic moment maps  (see e.g.~\cite{Andrianopoli:1996cm,deWit:2001bk})
\begin{align}\label{pot-mm}
  V^{P} = 2 \,\Bigl(g^{A \ov B} f_A^\Lambda f_{\ov B}^\Sigma -
  3 L^\Lambda \ov L{}^\Sigma\Bigr) P^x_\Lambda P^x_\Sigma\;.
\end{align}
These moment maps in turn are proportional to a covariant derivative on $\tilde
k^u_\Lambda$. However, as can be seen from~\eqref{eq:cov-dev-hypers},
the hypermultiplets are only gauged with the graviphoton $A_\mu^0$. Therefore $\tilde k^u_\Lambda = 0$ for $\Lambda \neq
0$ and their covariant derivative also vanishes, so $P^x_\Lambda = 0$  for $\Lambda \neq
0$. The only term in \eqref{pot-mm} that can contribute is  the term with $\Lambda=0$. We then utilize that the vector geometry is specified
by~\eqref {eq:special-prepotential}, from which one calculates $g^{A \ov B}
f_{\vphantom{\ov B}A}^0 f_{\ov B}^0 - 3 L^0 \ov L{}^0 = 0$. Combining these properties,
one finds that $V^P = 0$. This analogue of the
$\mathcal N=1$ no-scale property reduces the full scalar potential to~\eqref{eq:formal-potential}.

To see that~\eqref{eq:formal-potential} reproduces our scalar
potential, we 
 use $k^A_\Lambda \ov X{}^\Lambda = 2i M^A{}_B X^0 \phi^B$,
and as~\eqref{eq:cycl_condition} implies $\K_A M^A{}_B \phi^B = 0$, with the help of~\eqref{eq:special-kahler} we find
\begin{align}
  2\, {\rm e}^{\mathcal K} g_{AB}\, k^A_\Lambda k^{\ov B}_\Sigma \,\ov X{}^\Lambda
  X^\Sigma = -\frac 1 {4R^6}\, \K_{AB}
  M^A{}_C \phi^C M^B{}_D \phi^D\;.
\end{align}
Employing  the expressions for the covariant derivatives of the hyperscalars above, it is then
straight-forward to check that~\eqref{eq:formal-potential}
reproduces~\eqref{potential_4}.

%%%%%%%%%%%%%%%%%%%%%%%%%%%%%%%%%%%%%%%%%%%%%%%
%%%%%%%%%%%%%%%%%%%%%%%%%%%%%%%%%%%%%%%%%%%%%%%
%%%%%%%%%%%%%%%%%%%%%%%%%%%%%%%%%%%%%%%%%%%%%%%
%%%%%%%%%%%%%%%%%%%%%%%%%%%%%%%%%%%%%%%%%%%%%%%

\section{M-theory on twisted seven-manifolds}
\label{sec:direct_reduction}

The Scherk-Schwarz reduction described above,  yielding the gauged supergravity
La\-grang\-ian~\eqref{eq:n2lagr}, can also be obtained from a
compactification of eleven-dimensional supergravity on a seven-manifold. This point of view had also been taken 
in~\cite{Aharony:2008rx} for the vector multiplets. We will briefly review and extend this procedure in 
the present section to also include the hypermultiplet sector.

The seven-dimensional space  we are going to compactify on, denoted by $\mathcal Y$ in the following, is chosen as a fibration of a Calabi-Yau three-fold $\mathcal{X}$ over a circle $S^1$.
\eq{
  \label{compactf}
  \arraycolsep2pt
  \begin{array}{ccc}\mathcal{X} &\to & \mathcal{Y} \\
   && \downarrow \\[1mm]
   && S^1
   \end{array}
}
The coordinates of $\mathcal{X}$ will be
denoted by $y$ and the coordinate $z$ of the circle  is again normalized
such that $z\sim z+1$. At a particular reference point $z_0=0$, we choose a basis of harmonic two- and three-forms of the corresponding Calabi-Yau three-fold as in section \ref{sec::cy_geometry}. We then must indicate how this data changes when moving around the circle. 

In words, the difference to the point of view taken in section \ref{sec:scherk_schwarz} can be explained as follows: instead of specifying the $z$-dependence in the coefficient functions (i.e. the five-dimensional fields) as we do in the Scherk-Schwarz reduction, we can shift the $z$-dependence  from the fields into the basis of two- and three-forms of $\cal X$. This produces a  seven-dimensional manifold of the type \eqref{compactf}, which by construction is equivalent to the Scherk-Schwarz reduction. We now explain this in some more detail.
 
%%%%%%%%%%%%%%%%%%%%%%%%%%%%%%%%%%%%%%%%%%%%%%%
%%%%%%%%%%%%%%%%%%%%%%%%%%%%%%%%%%%%%%%%%%%%%%%

\subsubsection*{Cohomology}
Let us begin our discussion  with the cohomology of the compactification space $\mathcal Y$.
Analogous to the harmonic $(1,1)$-forms on $\mathcal{X}$ we introduce 
\eq{
  \label{basis_1-1_hat}
   \hat \omega_A(y,z)\;,\hspace{60pt} A=1,\ldots,h^{1,1}(\mathcal{X}) \;,
}
with $y$ denoting the coordinates on $\mathcal X$ and $z$ denoting the coordinate of the circle.
The dependence of $\hat\omega_A$ on $z$  is taken as
\begin{align}
\label{two-form_dep_01}
\hat  \omega_A(y,z) = \Bigl[ \exp \bigl(zM^T\bigr)  \Bigr]_A^{\;\;\;B} \;\omega_B \;,
\end{align}
where the exponential of the matrix $(M^T)_A{}^B$ is understood as a
matrix product and $\omega_B$ is a basis of harmonic $(1,1)$-forms on
the Calabi-Yau three-fold at a particular reference point $z_0=0$. The matrix $M^B{}_A$ is not arbitrary but, as explained in \cite{Aharony:2008rx}, has to satisfy the constraint shown in \eqref{eq:cycl_condition}.
Infinitesimally, the relation \eqref{two-form_dep_01} can be written as 
\begin{align}\label{eq:twisting-vectors}
  {\rm d} \hat \omega_A = \bigl(M^T\bigr)_A{}^B \hat \omega_B \wedge {\rm d}z \;,
\hspace{60pt} \hat\omega_B(y,0) = \omega_B \;,
\end{align}
so we see that in general the forms $\hat \omega_A$ are not closed. Their
non-closure will be the origin of the gaugings in the resulting
four-dimensional action.
The triple intersection numbers for the Calabi-Yau three-fold in the present context  are  given by
\eq{
  \label{def_tin_05}
  \hat{\mathcal{K}}_{ABC} \equiv \int_{\mathcal{Y}} \hat \omega_A\wedge \hat \omega_B\wedge
\hat  \omega_C  \wedge {\rm d}z  = \int_{\mathcal{X}} \omega_A\wedge\omega_B\wedge
  \omega_C = \K_{ABC}\;,
}
where the second equality follows by  using~\eqref{eq:cycl_condition}.

Analogous to the second cohomology, for the third cohomology  we introduce
\eq{
  \label{basis_3_real_hat}
  \bigl\{\hat \alpha_K(y,z) ,\hat \beta^L(y,z) \bigr\}\;,\hspace{40pt} K,L=0,\ldots ,h^{2,1}(\mathcal{X}) \;.
}
Their dependence on the coordinate $z$ of the circle is chosen as
\eq{
  \label{sym_struct_05}
  \binom{\hphantom{+}\hat \alpha(y,z)}{-\hat \beta(y,z)} = \Bigl[ \exp \bigl(zN^T\bigr) \Bigr] \binom{\hphantom{+}\alpha}
  {-\beta}\;,
}
where the matrix $N$ was defined in~\eqref{eq:definition-twist-n} and
proper contraction of indices is understood.  Furthermore,
$\{\alpha_K,\beta^K\}$ denotes the basis of harmonic three-forms on
the Calabi-Yau manifold at a particular reference point $z_0=0$, and the minus sign has been chosen to match the results from the previous section. Infinitesimally, we can express~\eqref{sym_struct_05} as\begin{align}\label{eq:twisting-hypers}
  {\rm d} \binom{\hphantom{-}\hat \alpha}{-\hat \beta} = - N^T\, \binom{\hphantom{-}\hat \alpha}{-\hat
    \beta} \wedge {\rm d} z\;,
   \hspace{40pt}
   \binom{\hphantom{-}\hat\alpha(y,0)}  {-\hat\beta(y,0)} = \binom{\hphantom{-}\alpha}{-\beta}  \;,
\end{align}
where proper contraction of indices is again understood.
Finally, using \eqref{sym_struct_01} and \eqref{sym_struct_05}, one can show that
\eq{
  \label{sym_struct_02}
  \int_{\X}\hat \alpha_K\wedge \hat \beta^L  = \delta_K{}^L \;,\hspace{35pt}
  \int_{\X}\hat \alpha_K\wedge \hat \alpha_L =0\;,\hspace{35pt}
  \int_{\X}\hat \beta^K\wedge \hat \beta^L  = 0 \;.
}

%%%%%%%%%%%%%%%%%%%%%%%%%%%%%%%%%%%%%%%%%%%%%%%
%%%%%%%%%%%%%%%%%%%%%%%%%%%%%%%%%%%%%%%%%%%%%%%

\subsubsection*{Dimensional reduction}

For  the dimensional reduction of the M-theory action \eqref{action_m_01} on the seven-manifold $\mathcal{Y}$ we make the following ansatz for the space-time metric
\eq{
  {\rm ds}_{11}^2 = {\rm e}^{\frac{4}{3}\phi} R^{-1} g_{\mu\nu}\, \d x^{\mu}
  \d x^{\nu} 
  + {\rm e}^{\frac{4}{3}\phi}\, R^2 \bigl( \d z -
  A^0 \bigr)^2 + G_{mn} \d y^m  \d y^n \;,
}
where $R$ is the radius of the circle satisfying
\eqref{relation_r_45}, $A^0$ denotes the graviphoton one-form and
$G_{mn}$ is the metric of the Calabi-Yau threefold, whose fluctuations depend on $\delta v^A$ and $z^r$.
For the three-form potential $\hat C_3$, we consider an ansatz similar to \cite{Aharony:2008rx} but are more specific about the sector corresponding to the hypermultiplets. In particular, we consider 
\eq{
  \label{expansion_11}
  \hat C_3 &= c_3 + B \wedge (dz- A^0) + (A^A - b^A A^0 )\wedge \hat
  \omega_A + b^A \hat \omega_A \wedge dz + C_3\;, \\
C_3 &= \sqrt{2}\,\xi^{K}\hat \alpha_{K} - \sqrt{2}\, \tilde  \xi_K \hat \beta^K\;,
}
where $c_3$ is a four-dimensional three-form, $B$ denotes a
four-dimensional two-form, $A^A$ are one-forms and $b^A$ as well as
$(\xi^K,\tilde\xi_K)$ are scalars in four dimensions. For the
corresponding field strength $\hat F_4 = \d \hat C_3$, employing \eqref{eq:twisting-vectors} as well as \eqref{eq:twisting-hypers}, one finds
\eq{
  \hat F_4 = \quad 
  \d c_3 + \d B\wedge & \bigl( \d z - A^0 \bigr)  - B\wedge F^0 + F^a \wedge \hat\omega_A
  - b^A F^0 \wedge \hat\omega_A \\
  + Db^A \wedge \hat\omega_A\wedge & \bigl( \d z - A^0 \bigr)
  + \sqrt{2} \, \left[ \d \binom{\xi}{\tilde\xi}^T - 
   \binom{\xi}{\tilde\xi}^T N^T \d z\right] \wedge \binom{\hphantom -\hat\alpha}{-\hat\beta} \;,
}
where $F^0$ and $F^A$ are defined in \eqref{def_vec_01}.
Using  the above ans\"atze in the  eleven-dimensional action~\eqref{action_m_01},
one can perform the dimensional reduction. However, to make contact with~\eqref{eq:n2lagr}, we have to  dualize  $B$ to a scalar $a$ and $c_3$ to a constant $e_0$, chosen to be zero.
A non-zero choice for  $e_0$ would correspond to a non-trivial $z$-dependence for the five-dimensional field $a$ in the Scherk-Schwarz reduction of section \ref{sec:scherk_schwarz}, which we did not consider.

Taking into account these points, we then recover the four-dimensional action \eqref{eq:n2lagr}, as we have checked explicitly.

%%%%%%%%%%%%%%%%%%%%%%%%%%%%%%%%%%%%%%%%%%%%%%%
%%%%%%%%%%%%%%%%%%%%%%%%%%%%%%%%%%%%%%%%%%%%%%%
%%%%%%%%%%%%%%%%%%%%%%%%%%%%%%%%%%%%%%%%%%%%%%%
%%%%%%%%%%%%%%%%%%%%%%%%%%%%%%%%%%%%%%%%%%%%%%%
%%%%%%%%%%%%%%%%%%%%%%%%%%%%%%%%%%%%%%%%%%%%%%%
%%%%%%%%%%%%%%%%%%%%%%%%%%%%%%%%%%%%%%%%%%%%%%%
%%%%%%%%%%%%%%%%%%%%%%%%%%%%%%%%%%%%%%%%%%%%%%%

\section{Truncation to $\mathcal N=1$ supersymmetry}

We now perform a truncation of the theory studied in  section \ref{sec:n2sugra} from $\mathcal{N}=2$ to $\mathcal N=1$ supersymmetry.
To motivate this truncation, we note that  M-theory compactifications on seven-manifolds of the form $\mathcal{X}\times S^1$ can be related to orientifold compactifications of type IIA string theory \cite{Kachru:2001je}. In particular, consider M-theory on 
\eq{
  \label{compatf_02}
  \frac{\;\mathcal{X}\times S^1\;}{( \ov\sigma,-1)} \;,
}
where $\ov\sigma$ is an anti-holomorphic involution acting on the Calabi-Yau three-fold $\mathcal{X}$ and where $(-1)$ acts on the circle coordinate as $z\to -z$.  Upon dimensionally reducing on $S^1$, the resulting theory is  type IIA string theory on 
\eq{
  \frac{\;\mathcal{X}\;}{\;(-1)^{F_L} \Omega \:\ov\sigma\;} \;,
}
where $F_L$ is the left-moving space-time fermion number and $\Omega$ is the  parity operator on the string world-sheet.
Motivated by this observation,  in the present work  we will impose a
truncation similar to \eqref{compatf_02}.

For later purpose, we also observe that $\ov\sigma$ being anti-holomorphic means that $\ov\sigma^{\,*}\Omega \sim \ov \Omega$, where $\Omega$ is the holomorphic three-form of the Calabi-Yau manifold and $\ov\sigma^{\,*}$ denotes the action of $\ov\sigma$ induced on the cohomology. Utilizing the relation 
\eq{
  \label{cy_relation_01}
  \Omega\wedge \ov \Omega \sim J\wedge J \wedge J \;,
}
and applying $\ov\sigma^{\,*}$ to both sides, 
we  infer that the K\"ahler form $J$  has to be odd under the anti-holomorphic involution $\ov\sigma^{\,*}$.

%%%%%%%%%%%%%%%%%%%%%%%%%%%%%%%%%%%%%%%%%%%%%%%
%%%%%%%%%%%%%%%%%%%%%%%%%%%%%%%%%%%%%%%%%%%%%%%
%%%%%%%%%%%%%%%%%%%%%%%%%%%%%%%%%%%%%%%%%%%%%%%
%%%%%%%%%%%%%%%%%%%%%%%%%%%%%%%%%%%%%%%%%%%%%%%

\subsection{Defining  the truncation}
\label{sec::def_trunc}

%%%%%%%%%%%%%%%%%%%%%%%%%%%%%%%%%%%%%%%%%%%%%%%
%%%%%%%%%%%%%%%%%%%%%%%%%%%%%%%%%%%%%%%%%%%%%%%

\subsubsection*{Cohomology}

To define our truncation, we first consider an involution $\ov \sigma$ acting on a Calabi-Yau three-fold $\X$. The action $\ov\sigma^{\:*}$ induced on the cohomology groups of $\X$  splits them into even and odd sub-spaces. In particular, the basis of harmonic $(1,1)$-forms introduced in \eqref{basis_1-1} can be separated as
\eq{
  \arraycolsep2pt
  \begin{array}{lclclcl}
  \ov\sigma^{\,*} \omega_{\alpha} &=& + \omega_{\alpha} \;, &\hspace{60pt}& 
    \alpha &=& 1,\ldots, h^{1,1}_{+} \;,\\[2mm]
  \ov\sigma^{\,*} \omega_{a} &=& - \omega_{a} \;,& \hspace{60pt}& a &=& 1,\ldots, h^{1,1}_{-} \;,
  \end{array}
}
where $h^{1,1}_++h^{1,1}_- = h^{1,1}$. Since the K\"ahler form is odd under $\ov\sigma^{\,*}$, also the volume form on $\X$ is odd. Thus,  some triple intersection numbers have to vanish which leads to
\begin{align}
  \label{truncation_348}
  \K_{\alpha \beta \gamma} = \K_{\alpha b c} = 0 \;,
  \hspace{60pt} \K_{\alpha b} = 0 \;,
  \hspace{60pt} \K_{\alpha} = 0\;.
\end{align}
For the basis of the third cohomology group of $\X$ introduced in \eqref{basis_3_real}, we similarly observe
\eq{
  \label{basis_3_trunc_01}
  \arraycolsep2pt
  \begin{array}{lclclcl}
  \ov\sigma^{\,*} \alpha_{k} &=& + \alpha_k \;, &\hspace{60pt}&
  \ov\sigma^{\,*} \beta^{k} &=& - \beta^k \;, \\[1mm]
  \ov\sigma^{\,*} \alpha_{\lambda} &=& - \alpha_{\lambda} \;, &&
  \ov\sigma^{\,*} \beta^{\lambda} &=& + \beta^{\lambda} \;, \\
  \end{array}
}
where the indices $k$ and $\lambda$ jointly range from $0$ to
$h^{2,1}$. For the period matrix $\mathcal{M}$ introduced in
equations \eqref{app_intersections}, from \eqref{basis_3_trunc_01} we then infer that
\eq{
  \label{truncation_11}
  {\rm Re}\,\mathcal{M}_{\kappa\lambda} = 0 \;,\hspace{40pt}
  {\rm Re}\,\mathcal{M}_{kl} = 0 \;,\hspace{40pt}
  {\rm Im}\,\mathcal{M}_{k\lambda} = {\rm Im}\,\mathcal{M}_{\lambda k} = 0 \;.
}

%%%%%%%%%%%%%%%%%%%%%%%%%%%%%%%%%%%%%%%%%%%%%%%
%%%%%%%%%%%%%%%%%%%%%%%%%%%%%%%%%%%%%%%%%%%%%%%

\subsubsection*{Truncation of vector multiplets}

Motivated by our discussion at the beginning of this section about ordinary M-theory compactifications, 
we will truncate our $\mathcal N=2$ supersymmetric theory by
\eq{
  \ov \Sigma = \bigl( \ov \sigma , -1 \bigr) \;,
}
where $\ov\sigma$ is the anti-holomorphic involution considered above and $(-1)$ acts on the circle coordinate as $z\to -z$. As noted below \eqref{cy_relation_01}, the K\"ahler form $J$ is odd under $\ov\sigma^{\,*}$, which we extend to 
\eq{
  \label{truncation_01}
  \ov \Sigma^{\,*}  J = - J \;.
}
In terms of the expansion $J= v^A(z)\, \omega_A$,\footnote{To keep our notation short, we suppress the dependence of the fields on $x^\mu$ but only indicate the dependence on the circle coordinate $z$.}  we find that
equation \eqref{truncation_01}, evaluated at $z=0$, yields
\eq{
  \label{truncation_07}
  v^{\alpha}(0 ) = 0 \hspace{40pt}
  \xrightarrow{\quad \eqref{app_nus}{\rm ~and~}\eqref{eq:redefinitions_d4}\quad}\hspace{40pt}
  \phi^{\alpha}(0)=0 \;.
}
For general values of $z$, we employ \eqref{def_twist_vec_finite} and \eqref{app_nus} to express $J$ as
\eq{
  \label{dependence_z_01}
  J(z) = v^a(0) \left[ {\rm e}^{z M^T }\right]_a^{\;\;\;B} \omega_B \;.
}
Inserting this expansion into~\eqref{truncation_01} leads to the constraint that $M^a{}_b=0$.

Next, concerning the vector fields $A^A$, we require that the  M-theory three-form  \eqref{ansatz_c3}  satisfies
\eq{
  \label{truncation_02}
  \ov \Sigma^{\,*}  \hat C_3 = + \hat C_3 \;.
}
In particular, the term involving the five-dimensional vector fields
$A^A_{(5)}$ has to be even under $\ov\Sigma^{\,*}$. Performing a
analysis similar  as for the K\"ahler form at $z=0$, and using equation \eqref{vec_5_4}, we obtain 
\eq{
  A^a_{(4)}(0) = 0 \;, \hspace{60pt}
  b^{\alpha}(0) = 0 \;.
}
Furthermore, requiring $A^A_{(5)} \wedge \omega_A$ to be even under
$\ov\Sigma^{\,*}$ for all values of $z$ and employing \eqref{def_twist_vec} implies that $M^{\alpha}{}_{\beta}=0$. We thus arrive at 
\begin{align}
  \label{truncation_08}
  M^A{}_B = \begin{pmatrix} 0 &  M^\alpha{}_b \\M^a{}_\beta & 0 \end{pmatrix}\;.
\end{align}

Finally, recalling the five-dimensional metric \eqref{metric_5_01} and requiring it to be invariant under the action \eqref{truncation_01}, we see that the graviphoton $A^0$ is projected out, that is
\eq{
  A^0 = 0 \;.
}

%%%%%%%%%%%%%%%%%%%%%%%%%%%%%%%%%%%%%%%%%%%%%%%
%%%%%%%%%%%%%%%%%%%%%%%%%%%%%%%%%%%%%%%%%%%%%%%

\subsubsection*{Truncation of hypermultiplets}

To define the truncation of the hypermultiplets, let us consider the action of
the anti-holomorphic involution  on the holomorphic three-form $\Omega$. Similarly as
in~\cite{Grimm:2004ua}, 
we write
\eq{
  \label{truncation_347}
   \ov\sigma^{\,*} \Omega = {\rm e}^{2 i \Theta}\, \ov \Omega \;,
}
where $\Theta$ is a constant phase. As for the K\"ahler form, we extend \eqref{truncation_347} to $\ov \Sigma$ in the following way
\eq{
   \label{truncation_04}
   \ov\Sigma^{\,*} \Omega  = {\rm e}^{2 i \Theta}\, \ov \Omega \;.
}
Employing then the expansion of  $\Omega$ given in \eqref{def_j_o}, at $z=0$ equation \eqref{truncation_04} implies that $\Im ( e^{-i \Theta} Z^k(0))=0$ and similar relations for $Z^{\lambda}$, $G_k$ and $G_{\lambda}$. However, for later purpose, let us introduce the compensator $C$ defined in terms of the
four-dimensional dilaton $\phi$ and the K\"ahler potential
\eqref{kaehler_pot_01} for the complex structure moduli 
\eq{
\label{eq:def_compensator}
  C \equiv {\rm e}^{-\phi} {\rm e}^{\K^{\rm cs}/2} {\rm e}^{-i \Theta}\;.
}
Noting that $\phi$ as well as $\K^{\rm cs}$ are invariant under $\ov\Sigma$, \eqref{truncation_04} can be brought into the form $\ov\Sigma^{\,*} (C\Omega) = \ov{C\Omega}$, whose implications at $z=0$ read
\eq{
\label{eq:trunc_cs}
  \arraycolsep2pt
  \begin{array}{lclclcl}
  \Im \bigl( \,C Z^k(0) \,\bigr) &=& 0 \;, &\hspace{60pt}
  \Re \bigl( \,C G_k(0) \,\bigr) &=& 0 \;, \\[2mm]
  \Re \bigl( \,C Z^{\lambda}(0) \,\bigr) &=& 0 \;, &\hspace{60pt}
  \Im \bigl( \,C G_{\lambda}(0) \,\bigr) &=& 0 \;.
  \end{array}
}
As carefully discussed in~\cite{Grimm:2004ua}, the  equations on
the left in \eqref{eq:trunc_cs} project out $h^{2,1}$ real scalars, corresponding to half of
the complex structure deformations. The set of equations on the right
should not be interpreted as further truncations, but as constraints
on the triple intersection numbers $d_{rst}$ in \eqref{prepot_01}.

Next, requiring $C_3$ in the M-theory three-form  \eqref{ansatz_c3} to be invariant under $\ov\Sigma^{\,*}$ leads to
\eq{
  \label{truncation_09}
   \xi^{\lambda}(0) = 0 \;,\hspace{60pt}
   \tilde\xi_k (0) = 0 \;.
}
To study  the five-dimensional three-form $\tilde c_3$ in \eqref{ansatz_c3}, we write
\eq{
  \tilde c_3 = \mathcal{C}_3 + \mathcal{C}_2\wedge dz \;,
}
where $\mathcal C_3$ and $\mathcal{C}_2$ respectively are three- and
two-forms in four dimensions. Since $\tilde c_3$ has to be even under
$\ov\Sigma^{\,*}$, we see that $\mathcal{C}_2$ is projected
out. Furthermore, $\mathcal{C}_3$ in four dimensions is dual to a constant $e_0$, which in the analysis of section \ref{sec:scherk_schwarz} and \ref{sec:direct_reduction}  we have chosen to be zero. Therefore, the contribution of $\tilde c_3$ in the truncated theory vanishes, which translates to
\eq{
  a = 0\;.
}
Combining then all these constraints, we see that $2 h^{2,1}$ out of the $4 (h^{2,1}
+ 1)$ original hypermultiplet scalars survive the truncation. We will later show that these
scalars form chiral multiplets and that their target space is K\"ahler.

Finally, in the above analysis we studied~\eqref{truncation_02} and~\eqref{truncation_04} at $z=0$. To
satisfy these constraints for all values of $z$, additional restrictions on the matrices $\Q$, $\R$, $\s$ and $\T$ introduced in \eqref{twist_3_01} arise. In particular, employing \eqref{twist_3_01} as well as \eqref{twist_3_05}, in a similar fashion as in \eqref{dependence_z_01} one obtains
\eq{
  \label{truncation_10}
  \arraycolsep2pt
  \begin{array}{lcllcllclcl}
  \Q^k{}_l &=& 0 \;,\hspace{40pt} & \Q^{\lambda}{}_{\rho} &=& 0 \;,\hspace{80pt}
    \R^{k \lambda} &=& \R^{\lambda k} &=& 0 \;, \\[1mm]
  \T_k{}^l &=& 0 \;,\hspace{40pt} & \T_{\lambda}{}^{\rho} &=& 0 \;,\hspace{80pt}
    \s_{k \lambda} &=& \s_{\lambda k} &=& 0 \;.
  \end{array}
}

%%%%%%%%%%%%%%%%%%%%%%%%%%%%%%%%%%%%%%%%%%%%%%%
%%%%%%%%%%%%%%%%%%%%%%%%%%%%%%%%%%%%%%%%%%%%%%%
%%%%%%%%%%%%%%%%%%%%%%%%%%%%%%%%%%%%%%%%%%%%%%%
%%%%%%%%%%%%%%%%%%%%%%%%%%%%%%%%%%%%%%%%%%%%%%%

\subsection{Performing the truncation}

After having specified the truncation of the fields appearing in the $\mathcal{N}=2$ theory (at the point $z_0=0$), we can now apply these results to \eqref{eq:n2lagr}. Recalling that this action  was obtained by evaluating all five-dimensional fields at a particular reference point $z_0=0$, and  employing the results from section \ref{sec::def_trunc}, we find
\begin{align}
  \nonumber
  \mathcal{S}_{(4)}^{\rm trunc.} = \int_{\mathbb{R}^{3,1}} \biggl[ \;\;\quad&\frac{1}{2}\: R_{(4)} \star_4 1 
   + \frac{1}{4} \, {\rm Im}\, \mathcal{N}_{ \alpha \beta} \, \d A^{ \alpha} \wedge\star_4 
   \d A^{\beta}
   +\frac{1}{4} \, {\rm Re}\, \mathcal{N}_{\alpha \beta} \, \d A^{\alpha} \wedge
   \d A^{\beta} \\
  \label{eq:n1lagr}
   -\,&g_{ab}\: Dt^a\wedge\star_4 D\ov t^b 
   - \frac{1}{6} \: A^\alpha M_\alpha{}^b \wedge A^\gamma \wedge \d
   A^\delta\, \K_{b \gamma \delta}
   \\
  \nonumber
   -\,&   G_{I \ov J}
\; {\rm d} M^I \wedge \star_4 {\rm d} \ov M^J - V^{\rm trunc.}  \quad\biggr] \;.
\end{align}

%%%%%%%%%%%%%%%%%%%%%%%%%%%%%%%%%%%%%%%%%%%%%%%
%%%%%%%%%%%%%%%%%%%%%%%%%%%%%%%%%%%%%%%%%%%%%%%

\subsubsection*{Kinetic terms}

In the truncated theory, the covariant derivative acting on the complexified K\"ahler moduli $t^a$ takes the form
\eq{
  \label{gauging_11}
  D t^a = \d t^a - M^a{}_{\beta} A^{\beta} \;.
}
The gauge kinetic function $f_{\alpha\beta}$ for the vector fields is found using the explicit formulas for the period matrix $\mathcal{N}$ given in \eqref{def_im_n}, and \eqref{truncation_348} as
\eq{
  \label{gauge_kin_fun}
  f_{\alpha\beta} = - i\,\overline{\mathcal{N}}_{\alpha\beta} = i\, \K_{\alpha\beta c}t^c \;,
}  
which is holomorphic in $t^a$, as required by $\mathcal
N=1$ supersymmetry. Note that the truncation splits the $\mathcal N=2$ vector multiplets into $\mathcal N=1$ vector and chiral multiplets with bosonic fields $A^{\alpha}$ and $t^a$, respectively.

Turning to the reduction of the 
hypermultiplets, since the graviphoton $A^0$ is projected out,  the hyperscalars become
uncharged. 
For the truncation of the hypermultiplets from $\mathcal{N}=2$ to
$\mathcal{N}=1$, we can thus refer to the existing literature. In particular, employing the results of appendix C in~\cite{Grimm:2004ua}, the kinetic terms for the hypermultiplet scalars are given by
\begin{align}
  \label{trunc_hyper_scalar}
  -\int_{\mathbb{R}^{3,1}} G_{I \ov J}\; {\rm d} M^I \wedge \star_4
  {\rm d} \overline M^{\ov J}\;,
\end{align}
where $M^I = \{ N^k, T_\lambda \}$ collectively denotes the chiral fields
\begin{align}
\begin{split}
  N^k = \frac 12 \xi^k + i\hspace{0.5pt} \Re (C Z^k)\;,\hspace{60pt}
  T_\lambda = i \tilde \xi_\lambda - 2\hspace{0.5pt} \Re (CG_\lambda)\;.
\end{split}
\end{align}
The metric $G_{I \ov J} = \partial_{\vphantom{\ov M^J}M^I} \partial_{\ov M^J} \K^{\rm Q}$ in \eqref{trunc_hyper_scalar} is 
K\"ahler, and the corresponding K\"ahler potential $\K^{\rm Q}$ is given by~\cite{Grimm:2004ua}
\begin{align}
  \label{kaehler_pot_q}
  \K^{\rm Q} = -2 \log \left[\, 2 \int_{\mathcal X} \Re \bigl(C \Omega\bigr) \wedge
  \star_6 \Re \bigl(C \Omega\bigr) \right]\;.
\end{align}

%%%%%%%%%%%%%%%%%%%%%%%%%%%%%%%%%%%%%%%%%%%%%%%
%%%%%%%%%%%%%%%%%%%%%%%%%%%%%%%%%%%%%%%%%%%%%%%

\subsubsection*{Potential}

Next, we consider the truncation of the scalar potential \eqref{potential_4} leading to $V^{\rm trunc.}$. For the scalars $\phi^A$  we employ \eqref{truncation_07} and \eqref{truncation_08} to find
\eq{
  \label{trunc_pot_10}
  V^{(1)} = -\,& \frac{1}{4 R^6}\: \bigl( M^{\alpha}{}_c\, \phi^c \bigr)\,\bigl(M^{\beta}{}_d \,\phi^d\bigr)
  \: \K_{\alpha\beta}  \;.
}

For the truncation of the terms involving  $\xi$ and
$\tilde\xi$ we use  \eqref{truncation_09},
\eqref{truncation_10} and \eqref{truncation_11}. These merely imply
that we have to restrict the index ranges of $\xi$ and
$\tilde \xi$ in~\eqref{potential_4}. For later convenience, we express this result as
\eq{
  \label{trunc_pot_11}
  V^{(2)} = - \frac{{\rm e}^{2\phi}}{2R^3} \:   \binom{\xi}{\tilde\xi}^T  N^T \, \Pi \,
    N\:
    \binom{\xi}{\tilde\xi} + \frac{{\rm e}^{4\phi}}{4R^3} \left[ \:
    \binom{\xi}{\tilde \xi}^TN^T \Delta \:\binom{\xi}{\tilde\xi} \:   \right]^2 
   \;,
}
where we have defined
\eq{
  \label{def_pi}
\Pi = \:\binom{\mathcal{M}}{-\mathds{1}}\:
    \bigl( {\rm Im}\,\mathcal{M}\bigr)^{-1}\,
    \binom{\ov{\mathcal{M}}}{-\mathds{1}}^T \;,\hspace{60pt}
  \Delta &= \begin{pmatrix}0 & 1 \\ -1 & 0 \end{pmatrix}\;.
}
As mentioned, these formulas are understood with the restrictions \eqref{truncation_09},
\eqref{truncation_10} and \eqref{truncation_11} applied.

To make the truncation of the potential for the complex structure moduli $z^r$ more feasible, we first 
define
\eq{
  \mathcal G_{LK} = 2 \,(\Im G)_{LK} -2 \, \frac{\,(\Im G)_{LN} \ov Z{}^N \,
    (\Im G)_{KM} Z^M}{ Z^N \,(\Im G)_{NM} \ov Z{}^M}\;,
}
with $G_{LK} =  \partial_{ Z^L} G_K$ and
$(Z^K,G_K)$  the holomorphic sections introduced in
equation \eqref{def_j_o}.
Recalling then \eqref{def_curl_n} as well as that $z^r=\frac{Z^r}{Z^0}$, we can write
\eq{
  \label{pot_trunc_3}
  V^{(3)} & =  \frac{1}{R^3}\: \mathcal{N}^r \ov{\mathcal{N}}^s \, G_{r\ov s}  \\
&= \frac{{\rm e}^{\K^{\rm cs}}}{R^3 |C|^2 } \, \Bigl( -CZ^K \T_K{}^L + CG_K \R^{KL} \Bigr)\, 
  \mathcal{G}_{LM}
  \Bigl( \Q^M{}_N \ov {CZ^N} + \R^{MN}\ov {CG_N} \Bigr) ,  
}
where the restrictions~\eqref{eq:trunc_cs} and \eqref{truncation_10} are understood. Note that to arrive at the second line in \eqref{pot_trunc_3}, we utilized $\mathcal G_{LK}\overline Z^K = 0$. The  compensator $C$ was introduced in equation~\eqref{eq:def_compensator}.

%%%%%%%%%%%%%%%%%%%%%%%%%%%%%%%%%%%%%%%%%%%%%%%
%%%%%%%%%%%%%%%%%%%%%%%%%%%%%%%%%%%%%%%%%%%%%%%
%%%%%%%%%%%%%%%%%%%%%%%%%%%%%%%%%%%%%%%%%%%%%%%
%%%%%%%%%%%%%%%%%%%%%%%%%%%%%%%%%%%%%%%%%%%%%%%

\subsection{Superpotential and D-terms}

We will now bring the potentials \eqref{trunc_pot_10}, \eqref{trunc_pot_11} and \eqref{pot_trunc_3}  into the standard form of $\mathcal N =1$ supergravity given by\,\footnote{Again, there is an overall
  factor 2 with respect to the standard literature; see footnote~\ref{footnote::factor2}.}
\begin{align}\label{eq:formal_n1}
  V = 2\, {\rm e}^{\K} \Bigl(G^{{\hat I} {\ov {\hat J}}} D_{\hat I} W D_{\ov {\hat J}}
 \ov W - 3|W|^2 \Bigr) +  \bigl[(\Re f)^{-1}\bigr]^{\alpha\beta} D_\alpha
 D_\beta = V_F + V_D\;,
\end{align}
where we use $\hat M^{\hat I} = \{ N^k, T_\lambda, t^a \}$ to label all chiral fields
in the theory. Here, the K\"ahler covariant derivative reads $D_{\hat I} W
= \partial_{\hat I} W + (\partial_{\hat I} \K) W$, $\Re f_{\alpha\beta}$ is the real part of
the gauge kinetic function \eqref{gauge_kin_fun} and $D_\alpha$ are the moment maps
associated with the gauging of the chiral multiplets.
The K\"ahler potential $\K$ in~\eqref{eq:formal_n1} is the sum of  \eqref{eq:special-kahler} subject to the truncation \eqref{truncation_07}, and $\K^Q$ given in \eqref{kaehler_pot_q},
\begin{align}\label{eq:full-kahler}
  \K = \K^{\rm vec} + \K^Q\;.
\end{align}

%%%%%%%%%%%%%%%%%%%%%%%%%%%%%%%%%%%%%%%%%%%%%%%
%%%%%%%%%%%%%%%%%%%%%%%%%%%%%%%%%%%%%%%%%%%%%%%

\subsubsection*{D-term potential}
A non-trivial D-term potential arises as some of the chiral fields are
gauged. In our case, as can be inferred from \eqref{gauging_11}, 
only the chiral fields $t^a$ originating from the
projection of the $\mathcal N=2$ vector multiplets are gauged. We will therefore
show that their potential term \eqref{trunc_pot_10} is given by the D-term
potential.

To find an expression for $D_\alpha$, we can use the truncation of the original moment maps $P_K$ on the special K\"ahler
space, given in~\cite{Andrianopoli:2004im}. We then obtain 
\begin{align}
  \label{d_term_01}
  D_\alpha = i \hspace{1pt}( \hspace{1pt}M^T\hspace{1pt})_\alpha{}^a\, \partial_{t^a} \K^{\rm vec} 
  = -\frac 1 {4\hspace{0.5pt}R^3} ( \hspace{1pt}M^T\hspace{1pt})_\alpha{}^a \,\K_a\;.
\end{align}
Noting that the Killing vectors after the truncation are given by $k^a_\alpha =
M^a{}_\alpha$ as well as that $\partial_{\ov b}  \partial_{\vphantom{\ov b}a} \K^{\rm vec} = g_{a \ov b}$, we see that the $D_\alpha$'s obey
\begin{align}
k^a_\alpha = -i g^{a\ov b} \partial_{\ov b} D_\alpha\;,
\end{align}
and therefore are moment maps for the Killing vectors $k^a_\alpha$.  
Contracting then equation~\eqref{eq:cycl_condition} with $\phi^b \phi^c$ and restricting
the index $A$ to $\alpha$, we find
$\K_a M^a{}_\alpha = -2 \K_{\alpha\beta} M^\beta{}_a \phi^a$, which allows us to bring \eqref{d_term_01} into the form
\begin{align}\label{D-term}
  D_\alpha = \frac 1 {2\hspace{0.5pt}R^3}\,  \K_{\alpha\beta} M^\beta{}_a \phi^a\;.
\end{align}
Employing finally the expression~\eqref{gauge_kin_fun} for the real part of the gauge kinetic function, that is $\Re f_{\alpha\beta} = -\K_{\alpha\beta}$, we arrive at
\begin{align}
  \label{pot_4_n1_01}
  V_D =  \bigl[(\Re f)^{-1}\bigr]^{\alpha\beta}  D_{\alpha} D_{\beta} 
   =-\frac{1}{4 R^6}\: \bigl( M^{\alpha}{}_c\, \phi^c \bigr)\,\bigl(M^{\beta}{}_d \,\phi^d\bigr)
  \: \K_{\alpha\beta}
  = V^{(1)}\;.
\end{align}
So indeed, in the truncated theory the potential term for the fields $\phi^a$ is a D-term potential and thus fits into the framework of $\mathcal N=1$ supersymmetry.

Furthermore, notice that the gauge group has become abelian, $G=U(1)^{h^{1,1}_+}$, since the Killing vectors are constant and hence commute. One can check explicitly that the action is gauge invariant, and in particular, the D-term in \eqref{D-term} is gauge invariant. Also, as we will analyze in section \ref{sec:vacua},  the gauge group can be broken further due to a Higgsing of the gauge fields.

%%%%%%%%%%%%%%%%%%%%%%%%%%%%%%%%%%%%%%%%%%%%%%%
%%%%%%%%%%%%%%%%%%%%%%%%%%%%%%%%%%%%%%%%%%%%%%%
%%%%%%%%%%%%%%%%%%%%%%%%%%%%%%%%%%%%%%%%%%%%%%%
%%%%%%%%%%%%%%%%%%%%%%%%%%%%%%%%%%%%%%%%%%%%%%%

\subsubsection*{F-term potential}

We now turn to the F-term potential. As mentioned above \eqref{trunc_hyper_scalar}, the chiral fields $(N^k, T_\lambda)$ are ungauged. Therefore,  in an $\mathcal N=1$ supersymmetric theory their
contribution to their scalar potential has to originate from a
superpotential $W$. We will now show that the potential $V^{(2)} + V^{(3)}$ for $(N^k, T_\lambda)$ can indeed be expressed in terms of
\begin{align}\label{superpot}
  W = \frac 12\: \mathcal U^T \Delta N \,\mathcal{U}\;,
\end{align}
where $\Delta$ was defined in \eqref{def_pi}, the twisting matrix $N$ had been introduced in \eqref{eq:definition-twist-n} and  the restrictions \eqref{truncation_10} are to be imposed. Furthermore, we have combined the chiral fields $N^k$ and $T_\lambda$ into the vector 
\begin{align}
  \label{def_vector_u}
  \mathcal{U} = \binom{2 i N^k}{T_\lambda} = \binom{i\, \xi^k - 2 \Re (CZ^k)}{i\, \tilde
    \xi_{\lambda} - 2 \Re(CG_{\lambda})} = i \,\mathcal{U}_I + \mathcal{U}_R\;.
\end{align}

To show that the superpotential \eqref{superpot} reproduces the scalar  potential $V^{(2)} + V^{(3)}$, we first note that
\eq{
  \partial_{t^a} W = 0 \;,\hspace{60pt}
  \partial_{\vphantom{\ov t^{\ov  b}}t^a} \K \, g^{a \ov b}\, \partial_{\ov t^{\ov  b}} \K = 3 \;,
}
where the K\"ahler potential $\K$ is given by \eqref{eq:full-kahler} and $g_{a \ov b}$ denotes the K\"ahler metric for the $t^a$.
This reduces $V_F$ in~\eqref{eq:formal_n1} to
\begin{align}
  \label{pot_4_n1_02}
  V_F = 2\, {\rm e}^{\K} \Bigl( \, G^{I \ov J} D_I W D_{\ov J} \ov W\, \Bigr)\,,
\end{align}
with $I$ labeling $(T_\lambda, N^k)$.
Next, we recall from~\cite{Grimm:2004ua}  the expressions for the corresponding inverse K\"ahler metric $G^{I \ov J}$ which are given by
\begin{align}
\label{metric_inverse_15}
\arraycolsep2pt
\begin{array}{lcll}
  \displaystyle G^{T_\kappa \ov T_\lambda} &=& \displaystyle -2 \,{\rm e}^{-2\phi} 
    & \displaystyle \Bigl[\Im \M + (\Re \M) (\Im \M)^{-1}
  (\Re \M)\Bigr]_{\kappa\lambda}\;,\\[2.5mm]
  \displaystyle  G^{T_\lambda \ov N^k} &=& \displaystyle -i\, {\rm e}^{-2\phi} 
    & \displaystyle \Bigl[(\Re \M) (\Im \M)^{-1}\Bigr]_\lambda^{\;\;\;k}\;,\\[2.5mm]
  \displaystyle G^{N^k \ov N^l} &=& \displaystyle -\frac 12\, {\rm e}^{-2\phi} 
    & \displaystyle \Bigl[ (\Im  \M)^{-1} \Bigr]^{kl}\;,
\end{array}
\end{align}
where in the present case~\eqref{truncation_11} implies that some entries of $\Re\, \M$ and
$\Im\, \M$ are vanishing. From \eqref{metric_inverse_15}, we can then compute the contractions
\begin{align}
  G^{N^k \ov M^{\ov I}} \partial_{\ov M^{\ov I}} \K=-(N^k - \ov N^k)\;,
  \hspace{40pt}
 G^{T_\lambda \ov M^{\ov I}} \partial_{\ov M^{\ov I}} \K=-(T_\lambda + \ov T_\lambda)\;.
\end{align}
Employing the above expressions as well as \eqref{eq:special-kahler} and \eqref{kaehler_pot_q}, one can bring equation \eqref{pot_4_n1_02} into the following form
\eq{
  V_F= \frac{1}{R^3}\:\frac{1}{\mathcal U_R^T \,\Pi\, \mathcal{U}_R} \:\biggl[ \qquad
  &\mathcal{U}^T_R N^T \Delta^T \left( \Pi^{-1} - \frac{\mathcal{U}_R \:\mathcal{U}_R^T}
    {\mathcal U_R^T \,\Pi\, \mathcal{U}_R} \right) \Delta \,N \, \mathcal{U}_R \\
  +\:&\mathcal{U}^T_I N^T \Delta^T \, \Pi^{-1} \, \Delta \,N \, \mathcal{U}_I 
  + \frac{\left( \mathcal U_I^T \Delta \, N\, \mathcal U_I \right)^2}
  {\mathcal U_R^T \,\Pi\, \mathcal{U}_R} \qquad \biggr] \;,
}
where $\mathcal U_R$ and $\mathcal U_I$ had been defined in \eqref{def_vector_u} and the matrices $\Pi$ as well as $\Delta$ had been introduced in equation \eqref{def_pi}. To proceed, we compute
\eq{
   \mathcal U_R^T \,\Pi\, \mathcal{U}_R = -4 \,\bigl| C\bigr|^2\, Z^K (\Im \mathcal M)_{KL}\, \ov Z{}^L
   = 2 \,{\rm e}^{-2\phi} \;,
}
and, by carefully taking \eqref{eq:trunc_cs} into account, one finds 
\eq{
    \Pi^{-1} - \alpha\:\frac{\mathcal{U}_R \:\mathcal{U}_R^T}{\mathcal U_R^T \,\Pi\, \mathcal{U}_R}
     = \binom{\mathds 1}{\mathcal M} \left(
     - (\Im \mathcal M)^{-1} + \alpha\:\frac{Z\, \ov Z^T}{ Z^T (\Im \mathcal M) \ov Z}
     \right)  \binom{\mathds 1}{\ov{\mathcal M}}^T \;,
}
where our case of interest is $\alpha=0$ and $\alpha=1$. With these relations, from the terms involving $\mathcal U_I$ one can now reproduce the potential $V^{(2)}$ for the fields $\xi$ and $\tilde \xi$. For the remaining terms, we note that the period matrix $\mathcal M$ can be expressed using the matrix $G_{LK} =  \partial_{ Z^L} G_K$ as follows
\eq{
  \mathcal M_{KL} = \ov G_{KL} + 2i\, \frac{ (\Im G)_{KM} Z^M \: Z^N(\Im G)_{NL} }{Z^T
  (\Im G) Z} \;.
}
Employing then the relation \eqref{cons_10}, one can bring the terms involving $\mathcal{U}_R$ into the form \eqref{pot_trunc_3}.

In conclusion, we have outlined how the superpotential \eqref{superpot}  indeed reproduces the scalar potential $V^{(2)}+V^{(3)}$ and thus fits into the framework of $\mathcal N=1$ supersymmetry.

%%%%%%%%%%%%%%%%%%%%%%%%%%%%%%%%%%%%%%%%%%%%%%%
%%%%%%%%%%%%%%%%%%%%%%%%%%%%%%%%%%%%%%%%%%%%%%%
%%%%%%%%%%%%%%%%%%%%%%%%%%%%%%%%%%%%%%%%%%%%%%%
%%%%%%%%%%%%%%%%%%%%%%%%%%%%%%%%%%%%%%%%%%%%%%%

\subsection{Connection to  manifolds with $G_2$ structure}

In this subsection, we indicate a connection of the truncated theory studied above to
compactifications of M-theory on seven-manifolds with $G_2$
structure. 

A manifold has $G_2$ structure if its structure group is
contained in $G_2$, and if it features a globally defined, $G_2$-invariant,
real and nowhere-vanishing three-form $\Phi$. Note that if $\Phi$ is in addition harmonic, the manifold has $G_2$ holonomy.
In our present setting, motivated by \cite{JoyceI,JoyceII,Hitchin:2000jd,Chiossi:2002tw,Grimm:2004ua,Micu:2006ey}, we define  $\Phi$ as
\begin{align}
  \label{def_m_phi}
  \Phi =  \sqrt{2}\, \Bigl( R\,\V^{-\frac{1}{3}}\, J \wedge {\rm d}z +  \sqrt{8}\, \Re (C \Omega) \Bigr)\;,
\end{align}
where we remind the reader that $R$ denotes the radius of the circle, $\V$ is the volume of the Calabi-Yau three-fold, $J$ denotes its K\"ahler form while $\Omega$ is the holomorphic three-form.
Following then for instance \cite{Micu:2006ey}, as ${\rm d}z$, $J$ and $\Omega$ are globally defined and
nowhere-vanishing, one can show that \eqref{def_m_phi} defines a $G_2$ structure
on the seven-manifold $\mathcal Y$.

Using \eqref{def_m_phi}, one can express  the K\"ahler potential and the superpotential  in the following way~\cite{Harvey:1999as,Hitchin:2000jd,Gutowski:2001fm,Beasley:2002db,Grimm:2004ua}
\begin{align}
  \label{g2_def}
  \mathcal K = -3 \ln \left( \frac 1 7 \int_{\mathcal Y} \Phi \wedge
  \star_7 \Phi \right),\hspace{30pt}
  W = \frac 1 8 \int_{\mathcal Y} \bigl( \sqrt{2}\, C_3 + i \Phi\bigr) \wedge {\rm d}_7 
  \bigl( \sqrt2\,C_3+ i \Phi\bigr)\;.
\end{align}
To verify that the formulas in \eqref{g2_def} indeed reproduce the K\"ahler potential and superpotential of our truncated theory, we first note that 
the sum of  \eqref{eq:special-kahler} and \eqref{kaehler_pot_q} can be brought into the form
\eq{
  \label{kaehlerg2}
  \K = - \log \,\bigl[ \,8    R^3 \,\bigr] 
  -2 \log \left[ 2\hspace{0.5pt}\V^{\frac 13} R^{-1} \int_{\mathcal Y} \Re \bigl(C \Omega\bigr) \wedge
  \star_7 \Re \bigl(C \Omega\bigr) \right]\;.
}
In the second term the integral is over the seven-manifold
$\mathcal Y$ and its prefactor arises from the $zz$-component of the
metric \eqref{metric_5_01} by taking into account the Weyl rescaling
mentioned above equation \eqref{app_action_09}. From \eqref{eq:def_compensator} and  \eqref{def_phi}, utilizing $\star_6\Re \Omega = \Im\Omega$, one also finds the relations 
\eq{
  2\hspace{0.5pt}\V^{\frac{1}{3}} R^{-1} \int_{\mathcal Y} \Re \bigl(C \Omega\bigr) \wedge
  \star_7 \Re \bigl(C \Omega\bigr) = 
  2 \int_{\mathcal X} \Re \bigl(C \Omega\bigr) \wedge
  \star_6 \Re \bigl(C \Omega\bigr)  = {\rm e}^{-2\phi} = \V \;,
}
which allow one to   reproduce \eqref{kaehlerg2} from
the K\"ahler potential in \eqref{g2_def}. 
Concerning the superpotential,  employing~\eqref{twist_3_01} as well as~\eqref{twist_3_05}, we can express  \eqref{superpot} in the following way
\begin{align}\label{superpot2}
  W = \frac 14 \int_{\mathcal Y}  \Omega_c \wedge {\rm
    d}_{7} \Omega_c \;,
    \hspace{60pt}\Omega_c = C_3 +  i\,\sqrt{8}\, \Re (C \Omega) \;,
\end{align}
where $\mathcal Y$ is the seven-dimensional space given by \eqref{compactf} and $C_3$, subject to the truncation \eqref{truncation_09}, was defined in \eqref{expansion_11}. 
One then shows that the superpotential in~\eqref{g2_def}
reproduces~\eqref{superpot2}.

We finally remark that in the literature on M-theory compactifications
on manifolds with $G_2$ structure, one usually does not obtain D-terms.  Studying this question in more detail would be an interesting extension of our work.

%%%%%%%%%%%%%%%%%%%%%%%%%%%%%%%%%%%%%%%%%%%%%%%
%%%%%%%%%%%%%%%%%%%%%%%%%%%%%%%%%%%%%%%%%%%%%%%
%%%%%%%%%%%%%%%%%%%%%%%%%%%%%%%%%%%%%%%%%%%%%%%
%%%%%%%%%%%%%%%%%%%%%%%%%%%%%%%%%%%%%%%%%%%%%%%
%%%%%%%%%%%%%%%%%%%%%%%%%%%%%%%%%%%%%%%%%%%%%%%
%%%%%%%%%%%%%%%%%%%%%%%%%%%%%%%%%%%%%%%%%%%%%%%
%%%%%%%%%%%%%%%%%%%%%%%%%%%%%%%%%%%%%%%%%%%%%%%
%%%%%%%%%%%%%%%%%%%%%%%%%%%%%%%%%%%%%%%%%%%%%%%
%%%%%%%%%%%%%%%%%%%%%%%%%%%%%%%%%%%%%%%%%%%%%%%
%%%%%%%%%%%%%%%%%%%%%%%%%%%%%%%%%%%%%%%%%%%%%%%

\section{Vacuum structure}
\label{sec:vacua}

%%%%%%%%%%%%%%%%%%%%%%%%%%%%%%%%%%%%%%%%%%%%%%%
%%%%%%%%%%%%%%%%%%%%%%%%%%%%%%%%%%%%%%%%%%%%%%%

\subsubsection*{The $\mathcal{N}=2$ theory}

Let us now briefly analyze the vacuum structure of the $\mathcal N=2$ theory derived in section \ref{sec:n2sugra}. In particular, to determine the minima of the potential \eqref{potential_4} we first compute
\eq{
  \phi^A\:\frac{\partial}{\partial \phi^A} \:V = - 3 V \;,
}
which means that the potential is a homogeneous function of degree three in the fields $\phi^A$. Thus, a necessary condition for a minimum is that the potential $V$ vanishes. Since $V$ shown in \eqref{potential_4} is a sum of semi-positive  terms, each of those has to vanish independently.
The non-degenerate solutions therefore are
\eq{
  &0 \overset{!}{=} \mathcal N^r =  b^r - \frac{2}{3} \,\beta\, z^r + B^r{}_s z^s -\frac{1}{2}\, R^r{}_{st}{}^v a_v 
  z^s z^t \;, \\[2mm]
  &0 \overset{!} = N\,\binom{\xi}{\tilde\xi} \;, \hspace{100pt}
  0 \overset{!}= M^A{}_B\,\phi^B  \;.
}
Notice that, from the last two equations, the flat directions of the potential \eqref{potential_4}  for $(\xi^K,\tilde\xi_K)$ and $\phi^A$ are counted by the number of zero eigenvalues of the twisting matrices $N$ and $M$. The eigenvectors of these matrices define a finite dimensional subspace, characterizing  the directions where moduli are stabilized. In turn, the directions orthogonal to this subspace correspond to the flat directions. We remark that the analysis for the complex structure moduli $z^r$ is slightly more involved.

Of course, there are also degenerate solutions which can lead to a vanishing potential. Recalling  \eqref{potential_4}, these include configurations such as $(\xi^k,\tilde\xi_\lambda)=0$, $\phi^A=0$, $\phi\to -\infty$, $R\to\infty$, or where the matrices $G_{r\ov s}$, $\mathcal{M}$ and $\K_{AB}$ have zero eigenvalues.

Furthermore, since some of the scalar fields of the theory are gauged, a mass term for the gauge fields $A^\Lambda$ can be generated. More concretely, 
the Lagrangian contains terms of the form
\eq{
  \int_{\mathbb{R}^{3,1}} \biggl[ \; M_{\Lambda\Sigma}\, A^\Lambda \wedge\star_4 A^\Sigma
\;\biggr]
  \;, \hspace{50pt} \Lambda,\Sigma = 0, \ldots, h^{1,1} \;,
}
with the mass matrix $M_{\Lambda \Sigma}$ given by
\eq{
  \renewcommand{\arraystretch}{1.6}
  \arraycolsep2pt
  \begin{array}{lcl@{\hspace{1pt}}l@{\hspace{1pt}}l}
  \displaystyle M_{AB} &=& \displaystyle - \bigl( &\displaystyle M^T g \,M \bigr)_{AB} 
    & \displaystyle \Bigl|_{\rm min.}\;, \\
  \displaystyle M_{0A} &=& \displaystyle + \bigl( &\displaystyle b^T M^T g\, M \bigr)_{A}
    &\displaystyle \Bigl|_{\rm min.} \;, \\
  \displaystyle M_{00} &=& \displaystyle - &\displaystyle b^T M^T g \,M b 
    &\displaystyle \Bigl|_{\rm min.}\;.
  \end{array}
}
Here $g=g_{AB}$ denotes the K\"ahler metric \eqref{metric_16}, $b^A=\Re\, ( t^A)$ and matrix products are understood. Note that $M_{00}$ contains an additional term proportional to the scalar potential, which however vanishes in the minimum.

%%%%%%%%%%%%%%%%%%%%%%%%%%%%%%%%%%%%%%%%%%%%%%%
%%%%%%%%%%%%%%%%%%%%%%%%%%%%%%%%%%%%%%%%%%%%%%%

\subsubsection*{The $\mathcal{N}=1$ theory}

To study the vacua of the truncated theory, we first recall the D- and F-term potential given in \eqref{pot_4_n1_01} and \eqref{pot_4_n1_02}
\eq{
   \label{pot_scalar_14}
 V = V_D + V_F =  
   -\frac{1}{4 R^6}\: \bigl( M^{\alpha}{}_c\, \phi^c \bigr)\,\bigl(M^{\beta}{}_d \,\phi^d\bigr)
   \: \K_{\alpha\beta}
 + 2 {\rm e}^{\K} \Bigl( \, G^{I \ov J} D_I W D_{\ov J} \ov W\, \Bigr) \;.
}  
Similarly as for the $\mathcal N=2$ case, a necessary condition for minima in the fields $\phi^a$ reads
\eq{
  0 \overset{!} = \phi^a\:\frac{\partial}{\partial \phi^a} \:V = - 3 V \;,
}
which, since \eqref{pot_scalar_14} is a sum of semi-positive definite terms, implies that $V_D=V_F=0$. The non-degenerate solution to $V_D=0$ is given by $D_{\alpha}=0$ which means
\eq{
   M^{\beta}{}_c \phi^c = 0 \;,
}
whereas the non-degenerate  solution to $V_F=0$ leads to  $F_I=0$. One configuration satisfying this constraint reads
\eq{
  N \,\mathcal U  =0 \;,
}
where the vector $\mathcal U$ was defined in \eqref{def_vector_u}. However, other solutions involving for instance $(\xi^k,\tilde\xi_\lambda)=0$ are also possible.

For the mass terms of the vector fields we recall that the graviphoton $A^0$ as well as the fields $A^a$ are  projected out. We are thus left with 
\eq{
  \int_{\mathbb{R}^{3,1}} \biggl[ \; M_{\alpha\beta}\, A^{\alpha}\wedge\star_4 A^{\beta}
  \;\biggr] \;,
}
where with the help of \eqref{truncation_08} the mass matrix $M_{\alpha\beta}$ is found to be
\eq{
  M_{\alpha\beta} =   - \bigl( M^T g\, M \bigr)_{\alpha\beta} \, \Bigl|_{\rm min.}\;,
}
and the metric $g_{ab}$ takes indices $a,b=1,\ldots,h^{1,1}_-$.

%%%%%%%%%%%%%%%%%%%%%%%%%%%%%%%%%%%%%%%%%%%%%%%
%%%%%%%%%%%%%%%%%%%%%%%%%%%%%%%%%%%%%%%%%%%%%%%
%%%%%%%%%%%%%%%%%%%%%%%%%%%%%%%%%%%%%%%%%%%%%%%
%%%%%%%%%%%%%%%%%%%%%%%%%%%%%%%%%%%%%%%%%%%%%%%
%%%%%%%%%%%%%%%%%%%%%%%%%%%%%%%%%%%%%%%%%%%%%%%

\section{Conclusions and outlook}

In this paper, we have performed a detailed analysis of Scherk-Schwarz reductions of M-theory down to four spacetime dimensions, including both the vector multiplet and hypermultiplet sectors. These reductions yield gauged ${\cal N}=2$ supergravities, with a potential for the scalar fields that we explicitly computed. We have focussed on the bosonic sector of the theory, though the fermions can be treated in a similar way, such as to preserve supersymmetry of the full Lagrangian. Our analysis here is an extension of previous results in the literature \cite{Andrianopoli:2004im,Aharony:2008rx}. 

\pagebreak[1]

Furthermore, we have defined a truncation from ${\cal N}=2$ to ${\cal N}=1$, inspired by the rules of orientifold projections in type IIA string theory. These models are determined by the K\"ahler potential for the chiral ${\cal N}=1$ multiplets, the superpotential, the gauge kinetic functions, and the D-terms, all of which we have explicitly computed. Our results show a close relation to compactifications of M-theory on manifolds with $G_2$ structure, which would be interesting to understand in further detail. 

The models we obtained are not of immediate phenomenological relevance. This is because not all moduli are stabilized, and we have not identified which of the vacua lead to supersymmetry breaking. However, the inclusion of quantum corrections, both perturbative and nonperturbative, could provide additional mechanisms to stabilize the yet unfixed moduli. Due to the presence of D-terms in our models, this might lead to metastable vacua with a positive cosmological constant that could be relevant e.g. for inflationary models. We leave this interesting possibility for future research.

%%%%%%%%%%%%%%%%%%%%%%%%%%%%%%%%%%%%%%%%%%%%%%%
%%%%%%%%%%%%%%%%%%%%%%%%%%%%%%%%%%%%%%%%%%%%%%%
%%%%%%%%%%%%%%%%%%%%%%%%%%%%%%%%%%%%%%%%%%%%%%%
%%%%%%%%%%%%%%%%%%%%%%%%%%%%%%%%%%%%%%%%%%%%%%%
%%%%%%%%%%%%%%%%%%%%%%%%%%%%%%%%%%%%%%%%%%%%%%%

\vskip1cm
\subsubsection*{Acknowledgements}

E. P. would like to thank the Kavli Institute for Theoretical Physics, Santa Barbara for hospitality during part of this work. This research was supported in part by the National Science Foundation under Grant No. NSF PHY05-51164. S.V. thanks the organizers of the  Simons Workshop on Mathematics and Physics 2010 at Stony Brook, for support during the final stages of this work.

%%%%%%%%%%%%%%%%%%%%%%%%%%%%%%%%%%%%%%%%%%%%%%%
%%%%%%%%%%%%%%%%%%%%%%%%%%%%%%%%%%%%%%%%%%%%%%%
%%%%%%%%%%%%%%%%%%%%%%%%%%%%%%%%%%%%%%%%%%%%%%%
%%%%%%%%%%%%%%%%%%%%%%%%%%%%%%%%%%%%%%%%%%%%%%%
%%%%%%%%%%%%%%%%%%%%%%%%%%%%%%%%%%%%%%%%%%%%%%%
%%%%%%%%%%%%%%%%%%%%%%%%%%%%%%%%%%%%%%%%%%%%%%%
%%%%%%%%%%%%%%%%%%%%%%%%%%%%%%%%%%%%%%%%%%%%%%%
%%%%%%%%%%%%%%%%%%%%%%%%%%%%%%%%%%%%%%%%%%%%%%%

\clearpage
\begin{appendix}
\section{Some details on the dimensional reduction to $\mathbf{D=5}$}
\label{app_dim_red}

Let us begin with the dimensional reduction of the eleven-dimensional Ricci scalar appearing in the action \eqref{action_m_01}. We first decompose (up to total derivatives)
\eq{
   \frac{1}{2} \int \hat R\star 1 = \frac{1}{2} \int d^{11} \hat x \,
   \sqrt{\hat G} \, \biggl[ \; R_{(5)} + R_{\X} 
   &- \frac{1}{4} \, 
  \bigl( G^{ab} \partial_{\tilde\mu} G_{bc} \bigr) \bigl( G^{cd} \partial^{\tilde\mu} G_{da} \bigr) \\
   &+\frac{1}{4} \, \bigl( G^{ab} \partial_{\tilde\mu} G_{ab}  \bigr) \bigl( G^{cd}\partial^{\tilde\mu}
    G_{cd} \bigr)  \;\biggr] ,
}
where $R_{(5)}$ denotes the Ricci scalar computed from the five-dimensional metric $\tilde g_{\tilde\mu\tilde\nu}$, $R_{\X}=0$ is the Ricci scalar of the Calabi-Yau manifold $\X$ and $\partial_{\tilde\mu}$ are derivatives with respect to the five-dimensional coordinates $\tilde x^{\tilde\mu}$.
We then split the Calabi-Yau metric $G_{mn}$ into a constant background part $\mathring G_{mn}$ and fluctuations around this background
\eq{
  G_{mn} = \mathring{G}_{mn} + \delta G_{mn} \;.
}
Following \cite{Strominger:1985ks,tian,Strominger:1990pd,Candelas:1990pi}, the fluctuations (in a complex basis with holomorphic indices $a,b$ and anti-holomorphic indices $\ov a, \ov b$) can be expressed as 
\eq{
  \label{app_b_expansion}
  \arraycolsep2pt
  \begin{array}{lcllcl}
  \delta G_{a\ov b}  &=& \displaystyle -i\, \delta v^A \,(\omega_A)_{a\ov b} \;,
  &A&=&1,\ldots,h^{1,1}\;, \\[3mm]
  \delta G_{ab} &=& \displaystyle \frac{\V}{\int_{\X} \Omega\wedge\ov\Omega}\: \ov z^r \, (\ov\chi_r)_{a\ov a\ov b} \,\Omega^{\ov a\ov b}{}_{b} \;,
  \hspace{50pt}
  &r&=&1,\ldots,h^{2,1}\;,
  \end{array}
}  
where $\delta v^A$ are  fluctuations around the background value $\mathring v^A$ of the expansion parameters  of the K\"ahler form given in \eqref{def_j_o}. In the following, these will be combined into 
\eq{
   v^A=\mathring v^A+\delta v^A \;.
}
Furthermore, $\chi_r$ denotes a basis of harmonic $(2,1)$-forms on $\X$, and the holomorphic three-form  $\Omega$  was introduced in \eqref{def_j_o}. The volume $\V$ of the Calabi-Yau threefold was defined in equation \eqref{def_vol}.
At lowest order in the fluctuations, $\chi_{r}$ in $\delta G_{ab}$ does not depend on the five-dimensional coordinates whereas  $z^r$ (as well as $\delta v^A$ in $\delta G_{a\ov b}$) are functions of $\tilde x^{\tilde\mu}$. We also note the relation
\eq{
  \mathring G^{a\ov b} (\omega_A)_{a\ov b} =  \frac{i}{2}\,\frac{\K_{ABC}v^Bv^C}{\V} \;,
}
and we define and compute
\eq{
  g_{(5)AB} \equiv \frac{1}{4\V} \int_{\X} \omega_A\wedge\star \omega_B = - \frac{1}{4\V} \left( \K_{ABC}v^C - 
  \frac{\K_{ACD}v^Cv^D \K_{BEF}v^Ev^F}{4\V} \right) ,
}
as well as
\eq{
  \label{app_metric_cs}
  G_{r\ov s} \equiv  - \frac{\int_{\X} \chi_r \wedge \ov \chi_s  }{\int_{\X} \Omega\wedge\ov\Omega} 
  \;,\hspace{60pt} r,s=1,\ldots,h^{2,1} \;.
}
Up to second order in the fluctuations $\delta G$, we then find
\begin{align}
   \frac{1}{2} \int_{\mathbb{R}^{4,1}\times\X} \hat R\star 1 = \int_{\mathbb{R}^{4,1}} \biggl[ \; \frac{\V}{2}\,R_{(5)} \star_5 1
   &-\V\, g_{(5)AB}  \d v^A \wedge\star_5 \d v^B 
   -\V\, G_{r\ov s} \d z^r \wedge\star_5 \d\ov z^{\ov s} \nonumber\\
   & + \frac{\V}{2}\, \d \log\V\wedge\star_5 \d\log\V\quad \biggr] \;.
\end{align}

Let us next turn to the kinetic term for the three-form potential $\hat C_3$.
Using the ansatz \eqref{ansatz_c3}, we  compute
\eq{
    \label{app_action_02}
    - \frac{1}{4} \int_{\mathbb{R}^{4,1}\times\X} &\hat F_4 \wedge \star \hat F_4 = 
    - \frac{1}{4} \int_{\mathbb{R}^{4,1}} \biggl[\;
    \V\, \d\tilde c_3\wedge\star_5 \d\tilde c_3 
    + 4\V \, g_{(5)AB} \, \d A^A\wedge\star_5 \d A^B \\[2mm]
    & -2\, \bigl( {\rm Im}\,\mathcal{M} \bigr)^{-1\, KL}
    \Bigl( \d\tilde\xi_K - \mathcal{M}_{KN}  \d\xi^N \Bigr)  \wedge\star_5
    \Bigl( \d\tilde\xi_L - \ov{\mathcal{M}}_{LM}  \d\xi^M \Bigr)  \; \biggr] \;.
}
Here, we have employed
the period matrix $\mathcal M_{KL}$ which satisfies \cite{Suzuki:1995rt,Ceresole:1995ca}
\eq{
  \label{app_intersections}
  \arraycolsep2pt
  \begin{array}{lcl}
  \displaystyle  \int_{\X} \alpha_K\wedge\star_6 \alpha_L &=&
    \displaystyle \Bigl[ - \bigl( {\rm Im}\,\mathcal{M} \bigr) - \bigl( {\rm Re}\,\mathcal{M} \bigr)
    \bigl( {\rm Im}\,\mathcal{M} \bigr)^{-1} \bigl( {\rm Re}\,\mathcal{M} \bigr) \Bigr]_{KL} \;,
    \\[5mm]
  \displaystyle \int_{\X} \alpha_K\wedge\star_6 \beta^L &=&
    \displaystyle \Bigl[ - \bigl( {\rm Re}\,\mathcal{M} \bigr)
    \bigl( {\rm Im}\,\mathcal{M} \bigr)^{-1}  \Bigr]_{K}^{\hspace{8pt}L} \;, \\[5mm]
  \displaystyle  \int_{\X} \beta^K\wedge\star_6 \beta^L &=&
    \displaystyle \Bigl[ -  \bigl( {\rm Im}\,\mathcal{M} \bigr)^{-1}  \Bigr]^{KL} \;,
  \end{array}
}  
with matrix products understood and $\{\alpha_K,\beta^L\}$ denoting the basis introduced in \eqref{basis_3_real}. For the topological term in the action \eqref{action_m_01} we compute (up to total derivatives)
\eq{
    \label{app_action_03}
  &-\frac{1}{12}\int_{\mathbb{R}^{4,1}\times\X}  \hat F_4\wedge \hat F_4 \wedge \hat C_3 \\
  =\:& -\frac{1}{12}\int_{\mathbb{R}^{4,1}} \biggl[ \; 6\, \d\tilde c_3 \wedge \Bigl( \xi^K\d\tilde\xi_K  
  - \tilde\xi_K\d\xi^K\Bigr) + \K_{ABC}\,\d A^A\wedge \d A^B\wedge A^C \;\biggr] \;.
}
To dualize $\tilde c_3$ to a scalar field, we introduce a Lagrange multiplier $a$ for $d\tilde c_3$ and add this term to the combined action \eqref{app_action_02} and \eqref{app_action_03}. After solving the equations of motion for $\tilde c_3$ and substituting them back into the action, the terms involving $\tilde c_3$ become
\eq{
& -\frac 14  \int_{\mathbb{R}^{4,1}} \V \,d \tilde c_3 \wedge \star_5 \d
\tilde c_3 +2 \d \tilde c_3 \wedge 
(\tilde \xi_K\d \xi^K  - \xi^K\d \tilde \xi_K ) +2 \d \tilde c_3
 \wedge \d a 
\\
=\:&
  -\frac{1}{4} \int_{\mathbb{R}^{4,1}} \frac{1}{\V}\, \Bigl( \d a + \xi^K\d\tilde\xi_K -
  \tilde\xi_K\d\xi^K 
    \Bigr) \wedge \star_5 \Bigl( \d a + \xi^L\d\tilde\xi_L- \tilde\xi_L\d\xi^L 
    \Bigr) \;.
}
Finally, we combine the above expressions and  perform a Weyl rescaling  $\tilde g_{\tilde\mu\tilde\nu}\to \V^{-\frac{2}{3}} \tilde g_{\tilde\mu\tilde\nu}$ of the five-dimensional metric to arrive at
\begin{align}
  \nonumber
  \mathcal{S}_{(5)} = \int_{\mathbb{R}^{4,1}} \biggl[ 
  \; &+\frac{1}{2}\,R_{(5)} \star_5 1 - \frac{1}{6}\, \d \log\V\wedge\star_5 \d\log\V
   - g_{(5)AB}  \d v^A \wedge\star_5 \d v^B \\
  \nonumber
   &    - G_{r\ov s} \d z^r \wedge\star_5 \d \ov z^{\ov s}
    - \V^{\frac{2}{3}} \, g_{(5)AB} \, \d A^A\wedge\star_5 \d A^B 
    \\
  \nonumber
    &  -\frac{1}{4\V^2} \, \Bigl( \d a + \xi^K\d\tilde\xi_K - \tilde\xi_K\d\xi^K  
     \Bigr) \wedge \star_5 \Bigl( \d a + \xi^L\d\tilde\xi_L - \tilde\xi_L\d\xi^L 
    \Bigr) \\
  \nonumber
    & +\frac{1}{2\V} \bigl( {\rm Im}\,\mathcal{M} \bigr)^{-1\, KL}
    \Bigl( \d\tilde\xi_K - \mathcal{M}_{KN}  \d\xi^N \Bigr)  \wedge\star_5
    \Bigl( \d\tilde\xi_L - \ov{\mathcal{M}}_{LM}  \d\xi^M \Bigr) \\
  &
    \label{app_action_09}
   -\frac{1}{12}\: \K_{ABC}\,\d A^A\wedge \d A^B\wedge A^C  
    \;\; \biggr] \;.
\end{align}
As it turns out, the field $\V$ belongs to a hypermultiplet and so \eqref{app_action_09}
contains terms mixing hyper- and vector multiplets. To make contact with the standard formulation of $\mathcal{N}=2$ supergravity in five dimensions, we introduce new fields
\eq{
  \label{app_nus}
  \nu^A = \V^{-\frac{1}{3}}\, v^A\;.
}
By definition, due to \eqref{def_vol}, these satisfy $\frac 1 6\,\K_{ABC}
  \nu^A\nu^B\nu^C = 1$ and so there are $h^{1,1}$ scalar fields $\nu^A$ subject to one
constraint, as well as the independent field $\V$. We then arrive at the following form of the five-dimensional action
\begin{align}
  \label{app_action_5}
  \nonumber
  \mathcal{S}_{(5)} = \int_{\mathbb{R}^{4,1}} \biggl[ 
  \; &+\frac{1}{2}\,R_{(5)} \star_5 1 - \frac{1}{4}\, \d \log\V \wedge\star_5 \d\log\V
   + \frac 14 \K_{ABC} \nu^C  \d\nu^A \wedge\star_5 \d \nu^B \\
  \nonumber
   &    + \frac 14 \Bigl(\K_{ABC}\nu^C -\frac 14 \K_{ACD} \nu^C \nu^D
   \K_{BEF} \nu^E \nu^F\Bigr) \, \d A^A\wedge\star_5 \d A^B 
    \\
    & -\frac{1}{12}\: \K_{ABC}\,\d A^A\wedge \d A^B\wedge A^C 
    - G_{r\ov s} \d z^r \wedge\star_5 \d \ov z^{\ov s}  \\
  \nonumber
    &  -\frac{1}{4\V^2} \, \Bigl( \d a +  \xi^K\d\tilde\xi_K  - \tilde\xi_K\d\xi^K  
   \Bigr) \wedge \star_5 \Bigl( \d a + \xi^L\d\tilde\xi_L - \tilde\xi_L\d\xi^L 
    \Bigr) \\
  \nonumber
    & +\frac{1}{2\V} \bigl( {\rm Im}\,\mathcal{M} \bigr)^{-1\, KL}
    \Bigl( \d\tilde\xi_K - \mathcal{M}_{KN}  \d\xi^N \Bigr)  \wedge\star_5
    \Bigl( \d\tilde\xi_L - \ov{\mathcal{M}}_{LM}  \d\xi^M \Bigr)   
    \;\; \biggr] \;.
\end{align}

%%%%%%%%%%%%%%%%%%%%%%%%%%%%%%%%%%%%%%%%%%%%%%%
%%%%%%%%%%%%%%%%%%%%%%%%%%%%%%%%%%%%%%%%%%%%%%%
%%%%%%%%%%%%%%%%%%%%%%%%%%%%%%%%%%%%%%%%%%%%%%%
%%%%%%%%%%%%%%%%%%%%%%%%%%%%%%%%%%%%%%%%%%%%%%%

\section{Some details on the dimensional reduction to $\mathbf{D=4}$}
\label{app_dim_red_4}

%%%%%%%%%%%%%%%%%%%%%%%%%%%%%%%%%%%%%%%%%%%%%%%
%%%%%%%%%%%%%%%%%%%%%%%%%%%%%%%%%%%%%%%%%%%%%%%

\subsubsection*{Computations}

To perform the dimensional reduction of the five-dimensional action \eqref{action_5} (which is the same as \eqref{app_action_5}), we note that the inverse of the metric \eqref{metric_5_01} reads
\eq{
  \tilde g^{\tilde \mu\tilde\nu} = \left( \begin{array}{cc} R\, g^{\mu\nu}  & 
  R \,A^{0\,\mu} \\  R \,A^{0\,\nu} & R^{-2} + R \,A^0_{\rho} A^{0\,\rho} \end{array} \right) \;,
}
where $A^{0\,\mu}$ is the graviphoton with indices raised by the
inverse of the four-dimensional metric $g_{\mu\nu}$. The determinant
of $\tilde g_{\tilde\mu\tilde\nu}$ is given by
\eq{
  \det g_{\tilde\mu\tilde\nu} = R^{-2} \det g_{\mu\nu} \;.
}  
For the five-dimensional Ricci scalar, we then find
\eq{
  \label{app_reduction_10}
  \int_{\mathbb{R}^{4,1}} \frac{1}{2}\,R_{(5)} \star_5 1 = 
  \int_{\mathbb{R}^{3,1}}  \biggl[  \:
  \frac{1}{2}\: R_{(4)}\star_4 1 &- \frac{3}{4} \d\log R \wedge\star_4 \d \log R 
  - \frac{R^3}{4} \, \d A^0\wedge\star_4 \d A^0 \biggr] \;.
}
Under the symmetries \eqref{symmetry_vector_5} discussed in
section~\ref{sect:symmetries}, due to equation \eqref{eq:cycl_condition}, the volume $\V$ is independent of $z$ and so
we have chosen $\partial_z \V=0$. Upon dimensional reduction, the
corresponding term in the action keeps the same form, i.e.
\eq{
   \int_{\mathbb{R}^{4,1}} \biggl[ \: - \frac{1}{4}\, \d \log\V \wedge\star_5 \d\log\V\: \biggr]
   =    \int_{\mathbb{R}^{3,1}} \biggl[\:  - \frac{1}{4}\, \d \log\V \wedge\star_4 \d\log\V \:\biggr] \;.
}
However, for the scalars $\nu^A$ there is a non-trivial dependence  on
the coordinate $z$ of the circle, which we have specified in equation \eqref{def_twist_vec}. This leads to 
\eq{
  \label{app_terms_03}
   &\int_{\mathbb{R}^{4,1}} \biggl[ 
  \:  \frac 14 \K_{ABC}\, \nu^C  \d\nu^A \wedge\star_5 \d \nu^B  \:\biggr] \\
  = \,&     \int_{\mathbb{R}^{3,1}} \biggl[ \: \frac 14 \,\K_{ABC} \nu^C  D \nu^A \wedge\star_4 D \nu^B
     + \frac{1}{4R^3} \,\K_{ABC} \nu^C \bigl( M^A{}_D \nu^D \bigr) \bigl(M^B{}_E \nu^E \bigr)
     \star_4 1\:\biggr] \;,
}
where we have defined
\eq{
  D \nu^A = \d \nu^A + A^0 M^A{}_B \nu^B \;.
}
The computation for the remaining five-dimensional scalar fields in
the action \eqref{action_5} is completely analogous. On the other hand, the reduction of the five-dimensional vector fields is non-trivial. In particular, using \eqref{def_twist_vec} and \eqref{vec_5_4}, for the kinetic term one finds
\eq{
   & \int_{\mathbb{R}^{4,1}} \biggl[ \:
   \frac 14 \Bigl(\K_{ABC}\nu^C -\frac 14\, \K_{ACD} \nu^C \nu^D
   \K_{BEF} \nu^E \nu^F\Bigr) \, \d A_{(5)}^A\wedge\star_5 \d A_{(5)}^B \\
   =\,& \int_{\mathbb{R}^{3,1}} \biggl[ \:    \frac 14 \Bigl(\K_{ABC}\nu^C -\frac 14\, \K_{ACD} \nu^C \nu^D
   \K_{BEF} \nu^E \nu^F\Bigr) \times \\
   & \hspace{150pt}\times 
    \Bigl(  R \, F_{(4)}^A \wedge \star_4 F_{(4)}^B + \frac{1}{R^2}\,Db^A\wedge\star_4
   Db^B \Bigr) \;,
}
with the definitions
\eq{
  F_{(4)}^A = \d A_{(4)}^A - M^A{}_B A_{(4)}^B\wedge A^0 \;,
  \hspace{40pt}
  Db^A = \d b^A - M^A{}_B \bigl( A_{(4)}^B - b^B A^0 \bigr) \;.
}
For the Chern-Simons term in the five-dimensional action \eqref{action_5}, employing the constraint \eqref{eq:cycl_condition}, we find in agreement with \cite{Aharony:2008rx}
\eq{
  \label{app_terms_01}
  \int_{\mathbb{R}^{4,1}} \biggl[\: &-\frac{1}{12}\: \K_{ABC}\,\d A_{(5)}^A\wedge \d A_{(5)}^B\wedge 
  A_{(5)}^C \:\biggr] \\
 =  \int_{\mathbb{R}^{3,1}} \biggl[\:
   & -\frac 1 6 \, \K_{ABC}\, F_{(4)}^A   \wedge M^B{}_D\, A^D_{(4)} \wedge A_{(4)}^C 
   - \frac 1 {4} \, \K_{ABC}b^C F_{(4)}^A \wedge F_{(4)}^B \\
    &+\frac{1}{6}  \, \K_{ABC} b^B b^C \d A^0 \wedge F_{(4)}^A 
    -\frac{1}{12} \, \K_{ABC} b^A b^B b^C \d A^0   \wedge \d A^0 \qquad\biggr] \;.
}

%%%%%%%%%%%%%%%%%%%%%%%%%%%%%%%%%%%%%%%%%%%%%%%
%%%%%%%%%%%%%%%%%%%%%%%%%%%%%%%%%%%%%%%%%%%%%%%

\subsubsection*{Standard form of $\mathcal{N}=2$ gauged supergravity}

Let us now bring the above results into the standard form of $\mathcal N=2$ gauged supergravity in four dimensions. However, for ease of notation we will drop all subscripts  indicating four-dimensional quantities since this will be clear from the context.
\begin{itemize}

\item The Einstein-Hilbert term shown in equation \eqref{app_reduction_10} is already in the standard  form.

\item Concerning the scalars $\nu^A$ and $b^A$, we first define  fields $\phi^A$ in the following way
\eq{
  \phi^A = R\,  \nu^A \;,\hspace{60pt}
  R^3 = \frac{1}{6}\: \K_{ABC} \phi^A\phi^B \phi^C \;,
}\pagebreak[2]
where we have included the constraint \eqref{restriction_nu} in terms of the $\phi^A$. 
Collecting then all kinetic terms involving $\phi^A$ and $b^A$ from above, we can express them as
\eq{
  \int_{\mathbb{R}^{3,1}} \biggl[ -g_{AB}\: Dt^A\wedge\star_4 D\ov t^B \biggr] \;,
}
where we employed the definitions \eqref{def_t_01} as well as \eqref{metric_16}.

\item For the four-dimensional vector fields $A^A$ and $A^0$, we first
  recall the definitions \eqref{def_vec_01} and \eqref{def_vec_02} for
  the combined field strengths and structure constants,
  respectively. Next, we note that the period matrix derived
  from~\eqref{eq:special-prepotential} reads
\eq{
  \label{def_im_n}
  \arraycolsep2pt
  \begin{array}{lcllcl}
  {\rm Im}\, \mathcal{N}_{AB} &=& -4\,R^3\, g_{AB} \;,\hspace{72pt}&
  {\rm Re}\, \mathcal{N}_{AB} &=& -\K_{ABC}b^C \;,  \\[1.8mm]
  {\rm Im}\, \mathcal{N}_{A0} &=& +4\,R^3\, g_{AB} b^B \;, &
  {\rm Re}\, \mathcal{N}_{A0} &=& +\frac{1}{2}\,\K_{ABC}b^B b^C \;,\\[0.8mm]  
  {\rm Im}\, \mathcal{N}_{00} &=& -R^3\,\Bigl( 1 + 4  g_{AB} b^A b^B \Bigr) \;, &
  {\rm Re}\, \mathcal{N}_{00} &=& -\frac{1}{3}\, \K_{ABC}b^Ab^B b^C \;.
  \end{array}
}  
With $\Lambda, \Sigma = 0, \ldots, h^{1,1}$, the kinetic and topological terms for the vector fields are then
expressed as
\eq{
  \label{app_terms_02}
  \int_{\mathbb{R}^{3,1}} \biggl[ \,
   + \frac{1}{4} \, {\rm Im}\, \mathcal{N}_{\Lambda\Sigma} \, F^\Lambda \wedge\star_4 
   F^\Sigma
   +\frac{1}{4} \, {\rm Re}\, \mathcal{N}_{\Lambda\Sigma} \, F^\Lambda \wedge
   F^\Sigma
   \, \biggr] \;.
}

\item In equation \eqref{app_terms_01}, there is one term not contained in \eqref{app_terms_02} which can be brought into the following form
\eq{
 \int_{\mathbb{R}^{3,1}} \biggl[ \,
  - \frac{1}{6} \: A^A M_A{}^B \wedge A^C \wedge \d A^D \K_{BCD} \,\biggr] \;.
}

\item For the hypermultiplets, we first note that the reduction from five to four dimensions is very similar to the one presented in \eqref{app_terms_03}. Defining then
\eq{
  \label{def_phi}
  \V = {\rm e}^{-2\phi} \;,
}
one arrives at the kinetic terms given in \eqref{result_hypers_01}.

\item Let us finally comment on the scalar potential. As one can see for
  instance from  \eqref{app_terms_03}, the non-trivial dependence of
  the scalar fields on the circle coordinate $z$  will lead to a scalar potential in four
  dimensions. Collecting these terms also for the remaining scalar fields, one arrives at the potential given in \eqref{potential_4}.

\end{itemize}

\end{appendix}

%%%%%%%%%%%%%%%%%%%%%%%%%%%%%%%%%%%%%%%%%%%%%%%
%%%%%%%%%%%%%%%%%%%%%%%%%%%%%%%%%%%%%%%%%%%%%%%
%%%%%%%%%%%%%%%%%%%%%%%%%%%%%%%%%%%%%%%%%%%%%%%
%%%%%%%%%%%%%%%%%%%%%%%%%%%%%%%%%%%%%%%%%%%%%%%
%%%%%%%%%%%%%%%%%%%%%%%%%%%%%%%%%%%%%%%%%%%%%%%
%%%%%%%%%%%%%%%%%%%%%%%%%%%%%%%%%%%%%%%%%%%%%%%
%%%%%%%%%%%%%%%%%%%%%%%%%%%%%%%%%%%%%%%%%%%%%%%
%%%%%%%%%%%%%%%%%%%%%%%%%%%%%%%%%%%%%%%%%%%%%%%

\clearpage
\nocite{*}
\bibliography{references}
\bibliographystyle{utphys}

%%%%%%%%%%%%%%%%%%%%%%%%%%%%%%%%%%%%%%%%%%%%%%%
%%%%%%%%%%%%%%%%%%%%%%%%%%%%%%%%%%%%%%%%%%%%%%%
%%%%%%%%%%%%%%%%%%%%%%%%%%%%%%%%%%%%%%%%%%%%%%%
%%%%%%%%%%%%%%%%%%%%%%%%%%%%%%%%%%%%%%%%%%%%%%%
%%%%%%%%%%%%%%%%%%%%%%%%%%%%%%%%%%%%%%%%%%%%%%%
%%%%%%%%%%%%%%%%%%%%%%%%%%%%%%%%%%%%%%%%%%%%%%%
%%%%%%%%%%%%%%%%%%%%%%%%%%%%%%%%%%%%%%%%%%%%%%%
%%%%%%%%%%%%%%%%%%%%%%%%%%%%%%%%%%%%%%%%%%%%%%%

\end{document}